\documentclass{elsart}
\usepackage{epsfig}   
\usepackage{epsf}  
\def\etal{{\it et al}}  
\begin{document}  
\rightline{HEPSU 99-5}
\rightline{Oct. 1999}
\vspace{2mm}
\begin{frontmatter}  
  
\title{Performance of the CLEO III LiF-TEA Ring Imaging  
Cherenkov Detector in a High Energy Muon Beam}  
  
\author{M. Artuso, R. Ayad, F. Azfar, A. Efimov, S. Kopp,}  
\author{R. Mountain, G. Majumder, S. Schuh, T. Skwarnicki,}  
\author{S. Stone, G. Viehhauser and J. C. Wang\thanksref{NSF}}  
\thanks[NSF]{Supported by the National Science Foundation}  
\address{Syracuse University,  Syracuse, NY 13244-1130}  
\author{  
T. Coan, V. Fadeyev, I. Volobouev, and J. Ye\thanksref{DOE}}  
\thanks[DOE]{Supported by Dept. of Energy}  
\address{Southern Methodist University, Dallas, TX 75275-0175}  
\author{  
S. Anderson, Y. Kubota, and A. Smith\thanksref{DOE}}  
\address{  
University of Minnesota, Minneapolis, MN 55455-0112}  
\author{E. Lipeles\thanksref{DOE}}  
\address{California Institute of Technology, Pasadena, CA 91125-0001}

\begin{abstract} 

The CLEO III Ring Imaging Cherenkov detector uses LiF radiators to  
generate Cherenkov photons which are then detected by proportional wire  
chambers using a mixture of CH$_4$ and TEA gases. The first two photon  
detector modules which were constructed, were taken to Fermilab and tested  
in a beam dump that provided high momentum muons. We report on results  
using both plane and ``sawtooth" shaped radiators. Specifically, we   
discuss the number of photoelectrons observed per ring and the angular  
resolution. The particle separation ability is shown to be sufficient  
for the physics of CLEO III.  
  
\end{abstract}

\end{frontmatter}  
%PACS numbers: 03.30+p, 07.85YK

Submitted to Nuclear Instruments and Methods A
\section{INTRODUCTION} 
  
The CLEO III detector is designed to study decays of $b$ and $c$ quarks,  
$\tau$ leptons and $\Upsilon$ mesons produced in $e^+e^-$ collisions  
near 10 GeV center-of-mass energy. The new detector is an upgraded  
version of CLEO II \cite{CLEOII}. It will contain a new four-layer silicon strip  
vertex detector, a new wire drift chamber and a particle identification system  
based on the detection of Cherenkov ring images.  
  
The main physics goals of CLEO III include studies of the CKM matrix, CP  
violation and rare decays of $B$ mesons. These studies require, in  
most cases, the separation of charged kaons from charged pions. Examples  
include studying the rare decay $\overline{B}^o\to \pi^+\pi^-$, where the  
background is from $\overline{B}^o\to K^-\pi^+$. Measuring the CP violation rate  
asymmetry in $\overline{B}^o\to K^-\pi^+$, versus ${B}^o\to K^+\pi^-$, 
where only the particle identification can distinguish between the two  
reactions. Another example is measuring the rate for $B\to \rho \gamma$, where  
the dominant reaction is $B\to K^* \gamma$.   
  
There are many other physics goals where the role of the particle  
identification is useful but not as crucial as those mentioned above. As an  
example, consider the measurement of the CKM element $|V_{cb}|$ which is thought  
to be best done using the reaction ${B^-}\to D^{*o} \ell^- \bar{\nu}$ at  
the kinematic point where the $B$ transforms to a $D^{*o}$ at rest in the   
$B$ rest frame. Here reducing the background in the $D^o$ reconstruction
 becomes important.   

\begin{figure}[hbt] 
\vspace{-7.5cm} 
\centerline{\epsfxsize 7.0in  \epsffile{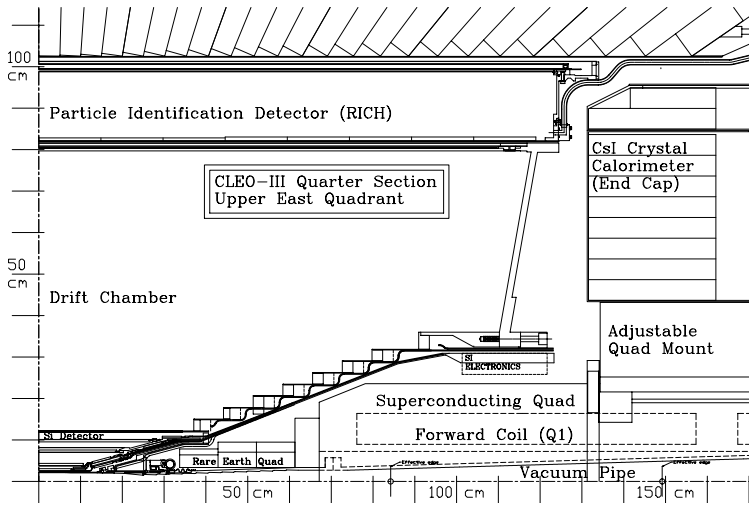}} 
\vspace{-8.2cm} 
\caption{\label{cleo} Schematic diagram of CLEO III (quarter section).} 
\end{figure}

CLEO III has approximate cylindrical symmetry with endcaps for some detector  
elements. Its outer shell consists of a muon detector, a super-conducting coil  
and an electromagnetic calorimeter that uses CsI crystals. These are the 
 same components as used in CLEO  
II. In order to achieve higher luminosity, the machine quadrupole magnets have been  
moved much closer to the collision point and necessitated the reconstruction 
of the detector inside the calorimeter. A sketch of CLEO III is shown in 
Fig.~\ref{cleo}  \cite{Artu98,Kopp96}. 
 
CLEO II produced many physics results, but was hampered by its limited 
charged-hadron identification capabilities. 
Design choices for particle identification were 
limited by radial space. The CsI calorimeter imposed a hard outer limit and the 
desire for maintaining excellent charged particle tracking imposed a  
lower limit, since at high momentum the error in momentum is 
proportional to the square of the track length. The particle identification 
system was allocated only 20 cm of radial space, and this limited the 
technology choices.  
 
The highest momentum particles from $B$ decays occur in the two-body decay 
$B\to \pi\pi$. Due to the small $B$ motion, the pions have a box shaped 
momentum distribution between 2.5 and 2.8 GeV/c. The goal of the particle 
identification system is to separate kaons from pions up to this momentum. 
It is useful to define the number of standard deviation separation as the 
difference between two measured quantities divided by the error on the 
measurement. For example 2$\sigma$ separation in Cherenkov angle would require 
that the difference between measured angles for pions and kaons divided by the 
average error on both measurement be equal to two. (In practice the errors 
on both measurements are nearly equal, especially when the difference in 
expected Cherenkov angle is small.) 
 
The drift chamber incorporates a measurement of specific ionization (dE/dx) 
which allows good kaon/pion separation below 800 MeV/c and useful, 
$\approx$1.8$\sigma$ separation, above 2.2 GeV/c to about 3 GeV/c. 
Thus an addition particle identification system should have at least   
3$\sigma$ separation at high momentum and at least 4$\sigma$ separation down 
to 800 MeV/c.

\section{GENERAL PRINCIPLES OF A PROXIMITY-FOCUSED RICH}   
%---------------------------------------------------------------------  
 
 %  
The CLEO III RICH detector   
consists of three components: radiator, expansion volume, and   
photon detector.    
No focusing is used; this is called ``proximity-focusing''   
\cite{t+j}.  %,js89},   
When an incident charged particle with sufficient velocity   
($\beta>1/n$) passes   
through a radiator medium,   
it emits photons   
at an angle $\Theta$ via the Cherenkov effect;   
some photons are internally reflected   
due to the large refractive index $n$ of the radiator, and some   
escape.    
These latter photons propagate in a transparent expansion volume,   
 %(e.g. pure N$_2$),   
sufficiently large to allow the Cherenkov cone   
to expand in size   
(as much as other spatial constraints allow).   
The position of the photoelectrons are determined by a two-dimensional array of
cathode-pads, in  
a photosensitive multi-wire chamber.

The resulting images are portions of conic sections,   
distorted by refraction and truncated by internal reflection   
at the boundaries of media with different optical densities.   
Thus, knowing the track parameters of the charged particle and the   
refractive index of the radiator, one can reconstruct the   
Cherenkov angle $\Theta=\cos^{-1}(1/n\beta)$ from the photoelectron
positions  
and extract the particle mass.    
 This elegant and compact approach was   
pioneered by the Fast-RICH Group \cite{Arno92,Guyo94,Segu94}.

In order to achieve efficient particle identification with low fake   
rates,   
we set as a design goal a system capable of   
$\pi/K$ separation with 4$\sigma$ significance   
($N_\sigma=\Delta\Theta/\sigma_\Theta$)   
at 2.65 GeV/$c$, the mean momentum for the pions from $B\to \pi\pi$
decays  
at a symmetric $e^+e^-$ collider.    
At this momentum, the $\pi/K$ Cherenkov angle difference   
$\Delta\Theta=14.4$ mr, which along with $1.8\sigma$ $dE/dx$   
identification 	%information   
from the central Drift Chamber,   
requires a Cherenkov angle resolution $\sigma_\Theta=4.0$ mr per   
track.   
 Since the estimated angular resolution per photon is 14 mr \cite{Efim95} using
 a 10 mm thick flat LiF radiator, we need approximately 12 photons
 to meet our goal.

\section{DESIGN} 
The 20 cm radial space limitation requires us to use a compact photon detector.
This is met by the use of Triethyleamine 
gas (TEA), mixed with methane. This mixture has a mean absorption length of 
$\approx$0.5 mm, but requires the use of light in the deep ultraviolet, between 
135-165 nm. Other choices such as TMAE gas or phototubes would have taken most of 
our allotted space, and thus were not possible. Other solutions such as DIRC \cite{DIRC} were not feasible for 
several reasons including  
the existing detector iron and the desire to limit the detector thickness to 
less than 12\% of a radiation length. 
 
\begin{figure}[hbt] 
\vspace{-1.5cm} 
\centerline{\epsfxsize 5.0in  \epsffile{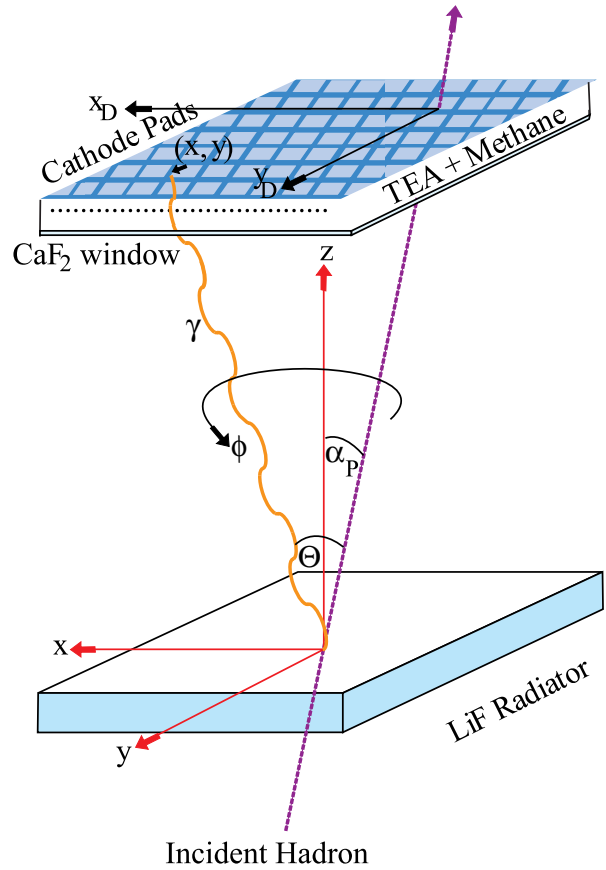}} 
\vspace{-3.2cm} 
\caption{\label{RICHscm} Schematic diagram of LiF-TEA RICH system.} 
\end{figure}

Transparent materials in this wavelength region are limited to  
flouride crystals. Thus the radiator is LiF and the chamber windows are 
CaF$_2$. 
The components of our system are illustrated in Fig.~\ref{RICHscm}. 
They consist of 
a LiF radiator,  in which UV photons are generated, an expansion region, and a 
photosensitive detector. The angle $\alpha_p$ is the polar angle of the 
incident particle with respect to the radiator normal, $\Theta$ is the 
Cherenkov angle and $\phi$ specifies the azimuthal angle of the Cherenkov 
photons. 

\begin{figure} [htb]
\vspace{4mm}  
\centerline{\epsfxsize 4.6in  \epsffile{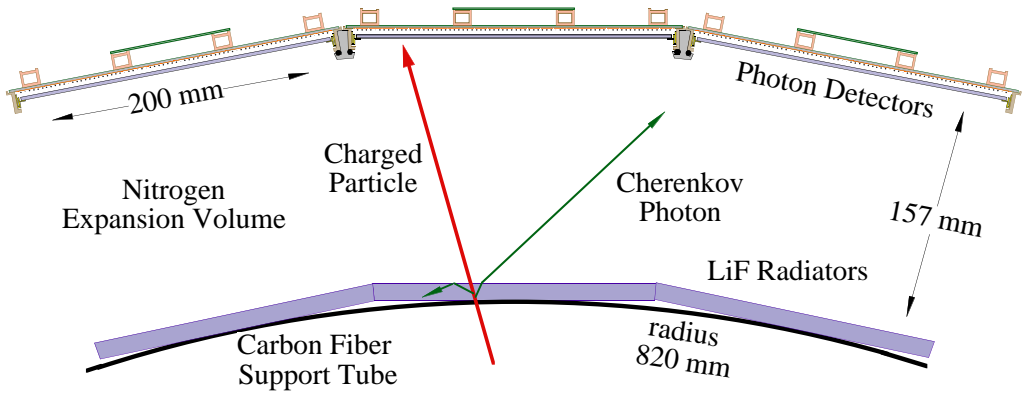}} 
\vspace{2cm} 
\caption{\label{rich2}A section of one tenth of the RICH detector as seen from the end.} 
\end{figure}

An end view of 1/10 of the detector is shown in Fig.~\ref{rich2}. The radiators
are glued to the outside of a carbon fiber cylinder of 82 cm radius. The
crystals (approximately 17.5 cm x 17 cm x 1 cm) are arranged in 30 rows
running parallel to the axis of the cylinder. Each row has 14 individual
crystals. The inner cylinder and 30 photon detector modules are attached to
end flanges and the space between them (the expansion volume) is filled
with ultraviolet light (UV) transparent N$_2$ gas.
 
\begin{figure} [htb]  
\centerline{\epsfxsize 4.5in \epsffile{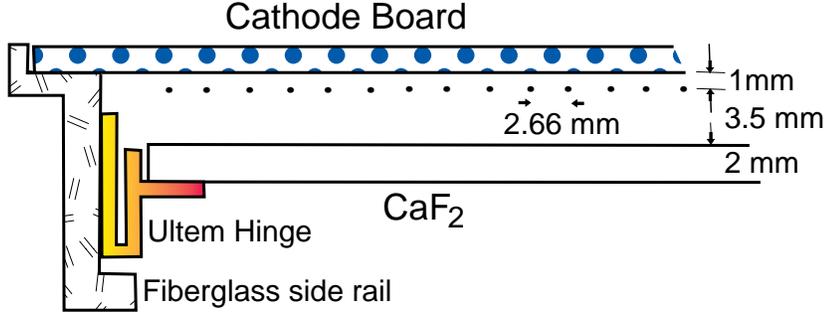}} 
\caption{\label{module_full}A cross-section of one RICH photon detection 
module.} 
\end{figure}

A sketch of a photon detector module is shown in Fig.~\ref{module_full}. 
The wires run the entire length of the detector, approximately 2.5 m. They are 
supported every 30 cm by a ceramic spacer. The CaF$_2$ window joints are  
placed 
directly on top of these spacers to minimize the loss of photon acceptance. 
Some parameters of the RICH detector are given in Table~\ref{params}.

\begin{table}[hbt] 
\begin{center}  
\caption{RICH photon detector parameters\label{params}} 
\begin{tabular}{l|l}\hline\hline 
Cathode pads & 24 in $x$ each 8  mm   and  320 in $y$ each 7.5  mm \\
Wires & 70  Gold plated tungsten with 3\% Rhenium, 20 $\mu$m in diameter\\
      & 2 guard wires on each side, gold plated tungsten 30 $\mu$m diameter\\
Wire spacings  & 2.66 mm between wires, 1 mm to cathode, 3.5 mm to 
CaF$_2$\\
Windows & 8 CaF$_2$, 2 mm thick,  30 cm x 19 cm (some half size)\\
Electrodes & Anode wires at +HV, 100 $\mu$m wide  silver traces on CaF$_2$ at -HV\\
\hline \end{tabular} \end{center}\vspace{2mm} \end{table}
 
%The container for the radiators and photon detectors consists of an inner 
%carbon fiber cylinder and an outer 30-sided polygon formed from the photon 
%detector modules. The inner and outer sectors are connected to end flanges, 
%forming a gas tight volume filled with N$_2$. The inner cylinder at a radius of 
%82 cm has individual LiF radiator tiles glued to the surface inside the N$_2$ 
%volume. There are 30 tiles segmented in azimuth and 14 rows along the length. 
%The outside cylinder is formed by 30 photon detector modules, each of which 
%spans the entire detector length of 280 cm.  

 \section{TEST SYSTEM ENCLOSURE}

Our goal to maximize the UV photon yield influences the design of the 
expansion volume for CLEO~III in three different ways. First, the
only mechanical connection between the inner cylinder of radiators 
and the outer
cylinder of photon detectors is located at the detector ends. This
avoids the obstruction of photon propagation by other structural elements 
inside the expansion volume. 
Second, supports between adjacent photon detectors and inside the photon 
detectors themselves are kept small to allow for maximum coverage with 
photon-sensitive surface. Chambers are separated only by $\sim5$~mm thick
aluminum rails, which mainly control the spacing between photon detectors. 
Structural stability is based on the stiffness of the cylinder formed by 
the photon detectors. Furthermore, we need to minimize the material
since the high quality CsI electromagnetic calorimeter is placed
behind the RICH. 

\begin{figure}[hbt]
   \centerline{\epsfxsize 4in \epsffile{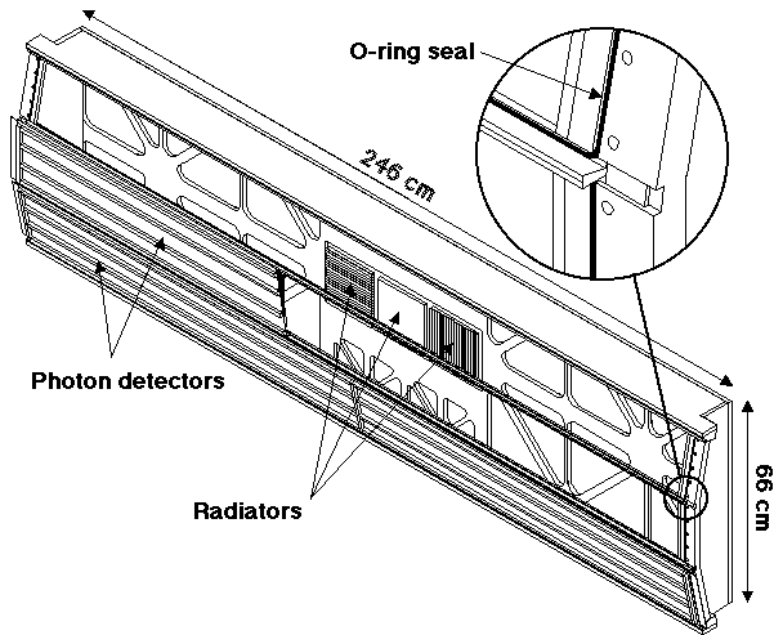} }
   \caption{Detector system enclosure used in the test beam.
Only part of the middle photon 
            detector is drawn to allow for better visibility.}
   \label{testbeambox}
\end{figure}

In order to satisfy these goals, we have a thin mechanical support system
that puts high demands on chamber stiffness. 
One of the goals for the test beam was to prove the mechanical soundness 
of the design chosen for CLEO~III. We therefore built an aluminum expansion 
volume box capable of supporting three photon detectors,
 that reproduces as closely as possible the final CLEO III photon detector
support. The box is shown in Fig.~\ref{testbeambox}. In particular, the middle
 photon detector 
was supported in exactly the same way as in the final design. In addition
the box provided mounting structures for the radiator crystals.

The final design calls for an expansion volume 
free of contaminants with a significant UV absorption in the sensitive 
wavelength range. Such contaminants would be oxygen, which has to be kept 
to a level $<10$ ppm to achieve 99\% transmission,  water or more complex, 
organic molecules.
To achieve the required purity we used a stainless steel gas system supplied 
with 99.99\% pure $N_2$. Purification 
included Oxisorb filters. 

A key element of the photon detector mounting scheme are 1/8'' diameter 
O-rings in the 
support structure running around the perimeter of each photon detector. They 
provide a gas tight separation between the expansion volume and 
the outside world while allowing for the removal of photon detectors if
access is required. 

\begin{figure}[hbt]
   \centerline{\epsfxsize 4in \epsffile{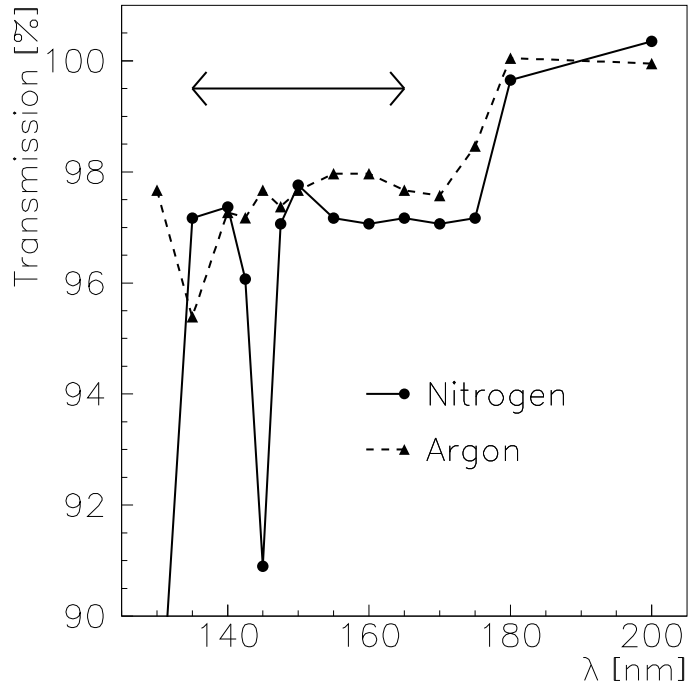} }
   \caption{Test beam expansion volume UV transparency for different 
            reference gases. The arrow shows the sensitive wavelength
            region for $\mathrm{CH}_4/\mathrm{TEA}$.}
   \label{transmission}
\end{figure}

The expansion 
volume transparency for UV photons was monitored by a system consisting of 
a deuterium lamp, a monochrometer, a gas volume containing the gas sample
and a photomultiplier tube. The transmission measurements were normalized 
to different reference gases of research grade quality.
Fig.~\ref{transmission} shows the results for the transparency of nitrogen 
after passage through the test beam expansion volume box. We observed an
absorption line at 145 nm. Its origin is not in contamination inside the 
expansion volume as the dip vanishes when the box is flushed with 99.98\%
pure Argon. This line and the one near 136 nm appear to be caused by
molecular transitions in the $N_2$ \cite{oldwilk}.

\section{RADIATORS} 
\subsection{Introduction} 
If a track is incident normal to a LiF radiator no light is emitted from
the radiator in the 
frequency range that can be detected by TEA, due to total internal reflection. 
A full system for CLEO, therefore, would require that the radiators be tilted 
until the track angles are about 20$^o$. This causes several serious 
problems. One is the difficulty in accurate radiator mounting. More 
importantly, some of the light from one radiator must traverse through another 
radiator resulting in losses and reconstruction problems. To eliminate 
these problems we invented  a novel radiator geometry. The light 
emitting surface of the radiator is cut with a profile  resembling 
the teeth of a saw, the ``sawtooth radiator" 
\cite{alex}.  The major advantage of this configuration is that  it reduces the 
losses of photons due to total internal reflection at the interface between the 
radiator and the expansion gap. It also turns out that the number of 
reconstructed photoelectrons is greatly improved and the  angular resolution per 
photoelectron is greatly decreased. 
 
\begin{figure} [htb] 
%\vspace{-2cm} 
\centerline{\epsfxsize 4.5in  \epsffile{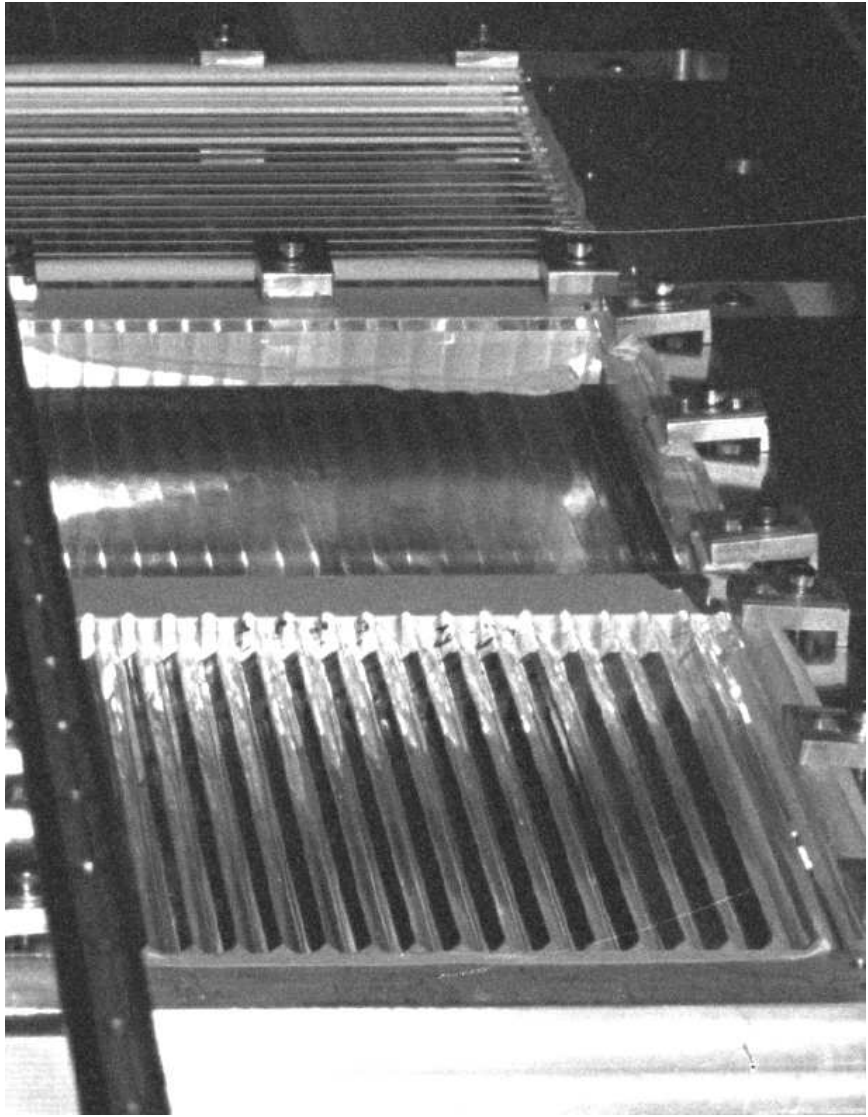}} 
\vspace{-.3cm} 
\caption{\label{saw_photo}Two sawtooth radiators separated by a plane radiator. The  
teeth are aligned perpendicular to one other in the test beam setup.} 
\end{figure}

A  profile with  teeth about 4 mm deep in a plate of 12 mm  thickness is 
a good compromise between uniformity in light output, and cost.  A  detailed  
simulation has shown that a tooth angle of 42$^\circ$ is close to optimal and 
technically feasible. (The tooth angle is defined by the intersection of
a plane parallel to the surface of a tooth with a plane parallel to the
base of the radiator.)   
There are several problems inherent in producing such radiators, including the 
ability to precisely cut the LiF without cleaving the material and the ability 
to polish the surface yielding good ultraviolet light transmission. 
Furthermore, the production time and cost need to be reasonable.  We 
worked with OPTOVAC in North  Brookfield, Mass. to produce full sized 
radiators, with dimensions 17.5 cm x 17.0 cm x 1.2 cm. The two sawtooth radiators separated by a 
plane radiator as used in the test beam are shown in Fig.~\ref{saw_photo}. 

\subsection{Shapes of Sawtooth Radiators}

\begin{figure}[htb] 
\vspace{.5cm} 
\centerline{\epsfig{figure=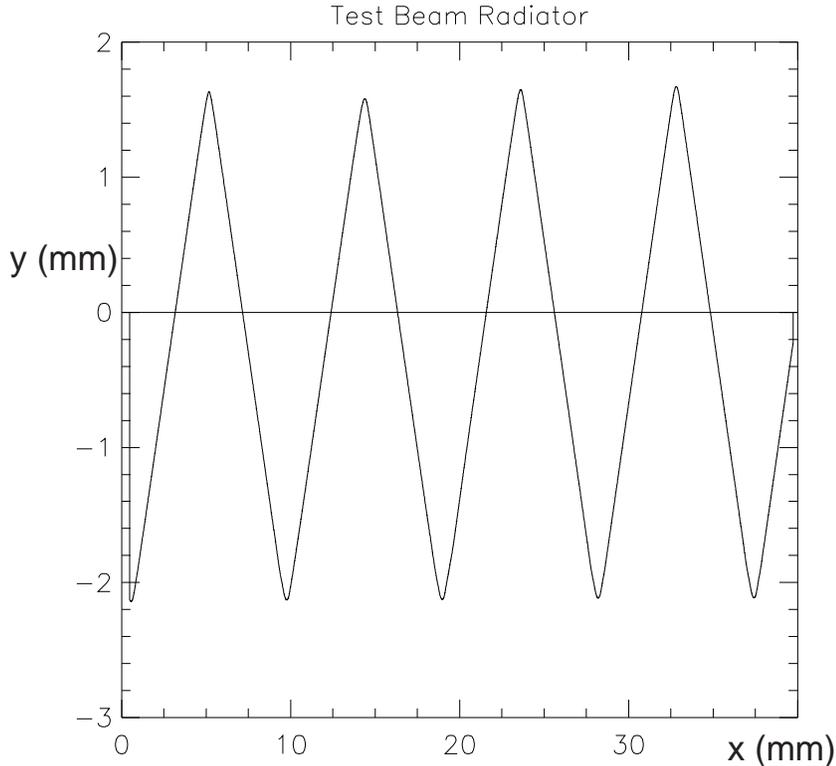,height=4in}} 
\vspace{-.4cm} 
\caption{\label{testbeam}The profile of the test beam radiator.} 
\end{figure}

We measured the shapes of two sawtooth radiators, one used in the test beam and another more advanced sample used in the final detector. 
A plot of measurements of the test 
beam piece is shown in Fig.~\ref{testbeam}. There are 40,000 points and the 
vertical precision is good to approximately 50 nm. Such accuracy is certainly 
overkill, but was provided by the Form-Talleysurf machine that was available at the Center of Optics 
Manufacture, University of Rochester. 
 
\begin{figure}[htb] 
%\vspace{-2.2cm} 
\centerline{\epsfig{figure=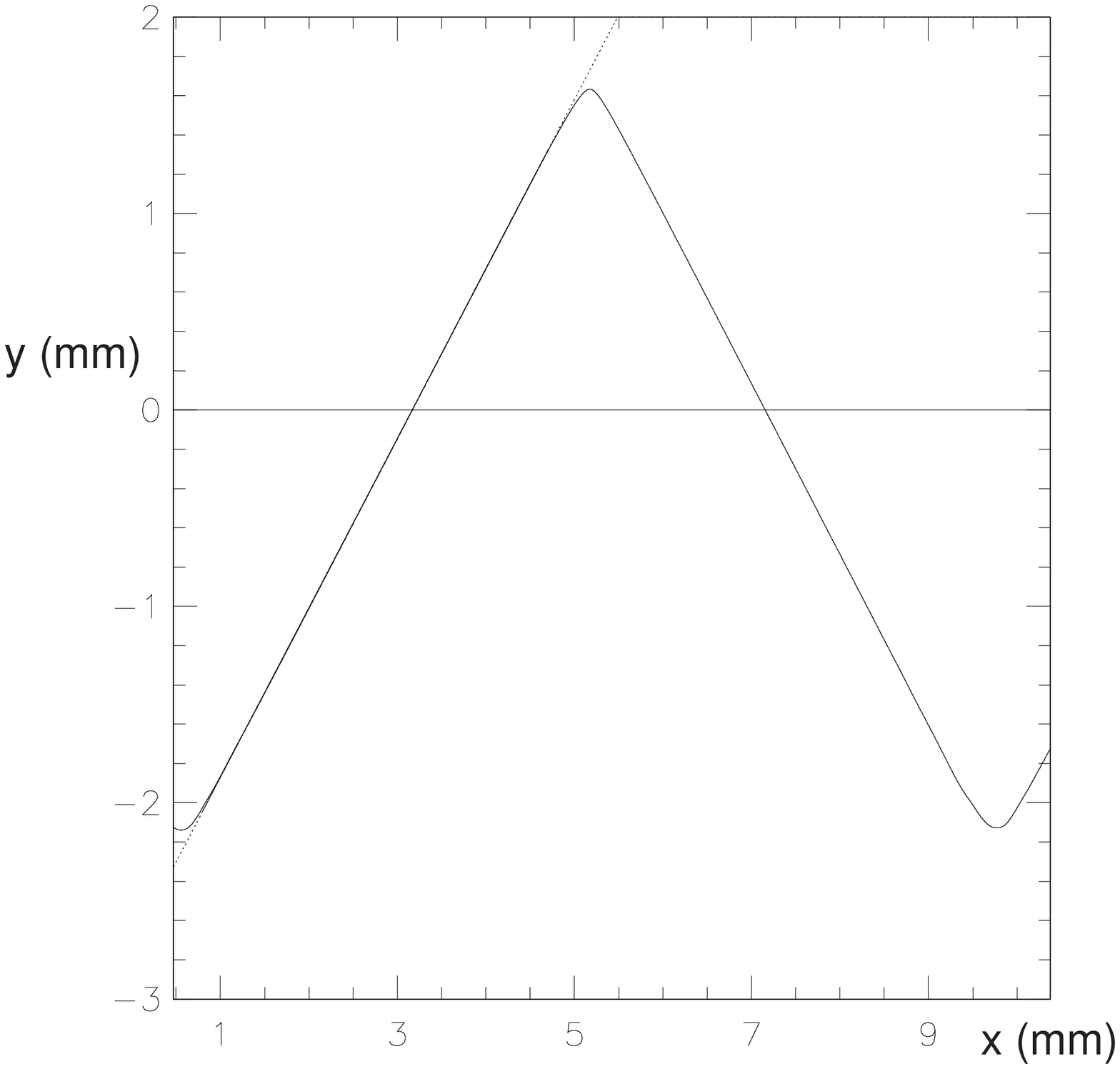,width=2.8in},
\epsfig{figure=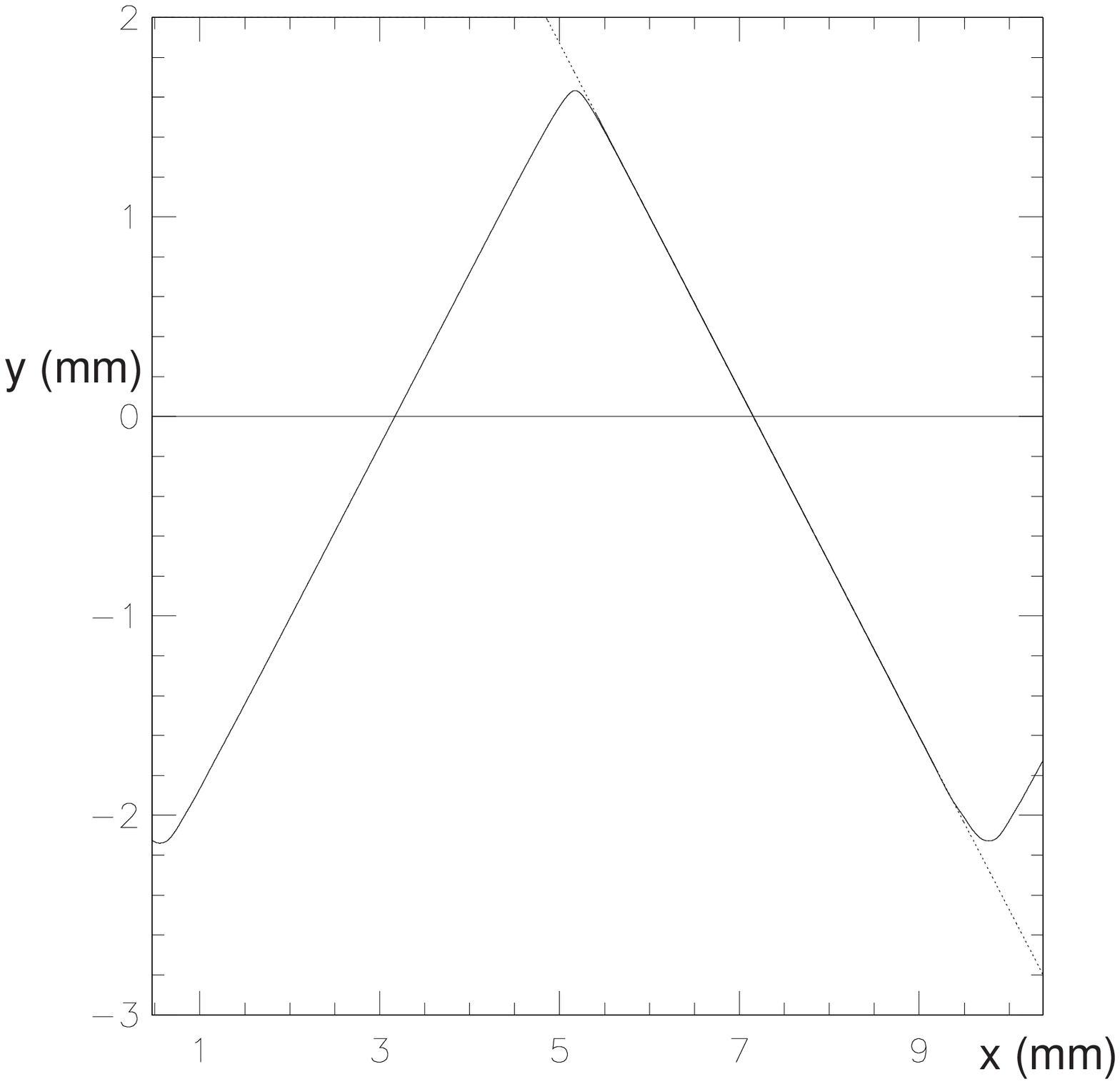,width=2.8in}} 
\vspace{-.1cm} 
\caption{\label{tb_1}Linear fits to one tooth of the test beam piece. The left 
side has a slope of 0.86298$\pm$0.00002, while the right side slope is
-0.86849$\pm$0.00002.} 
\end{figure}

Each tooth edge 
was fit to a straight line. An example is given in Fig.~\ref{tb_1}.
The angles can be changed slightly by changing the range of fit, but not 
significantly as will be demonstrated later. Fitting both sides of four teeth  
on each of the two radiators gave the measurements listed in  
Table~\ref{teethang}. 
 
\begin{table}[hbt] 
\begin{center}  
\caption{Angles of the various teeth (degrees)\label{teethang}} 
\begin{tabular}{lcccc}\hline\hline 
\multicolumn{1}{l}{tooth} & \multicolumn{2}{c}{Test Beam Radiator} 
&\multicolumn{2}{c}{New Radiator}\\\hline 
Face &Left Side & Right Side & Left Side & Right Side\\\hline 
1 & 40.794 & -40.974 & 41.609 & -41.612 \\ 
2 & 40.629 & -40.820 & 41.607 & -41.507 \\ 
3 & 41.088 & -41.073 & 41.635 & -41.526 \\   
4 & 41.077 & -41.079 & 41.586 & -41.601 \\ \hline 
avg & 40.90$\pm$0.11 & -40.99$\pm$0.06 & 41.61$\pm$0.01 &  -41.56$\pm$0.01 \\ 
rms (mr)& 3.9 & 2.1 & 0.4 & 0.4 \\ 
 \hline \end{tabular} \end{center} \end{table} 
 
For both radiators the left side and right side give equivalent values. 
These then can be averaged to give sawtooth angles of 40.90$\pm$0.07$^{\circ}$  
for the test beam radiator and 41.585$\pm$0.017$^{\circ}$ for the new radiator. 
 
The r.m.s. deviations are almost an order of magnitude larger for the test beam with  
respect to the new radiator, being on the order of 4 mr and 0.4 mr, respectively.
Although the 4 mr variation appears on first sight to be large enough to cause
a significant degradation in the resolution, our simulation shows that it 
adds a contribution of only 1 mr in quadrature to the final Cherenkov 
resolution per track. 
Variations in tooth angles in the new radiators will add only a negligible,
$\approx$0.1 mr variation in quadrature to the Cherenkov track angle
resolution.

\begin{figure}[htb] 
%\vspace{.9cm} 
\centerline{\epsfig{figure=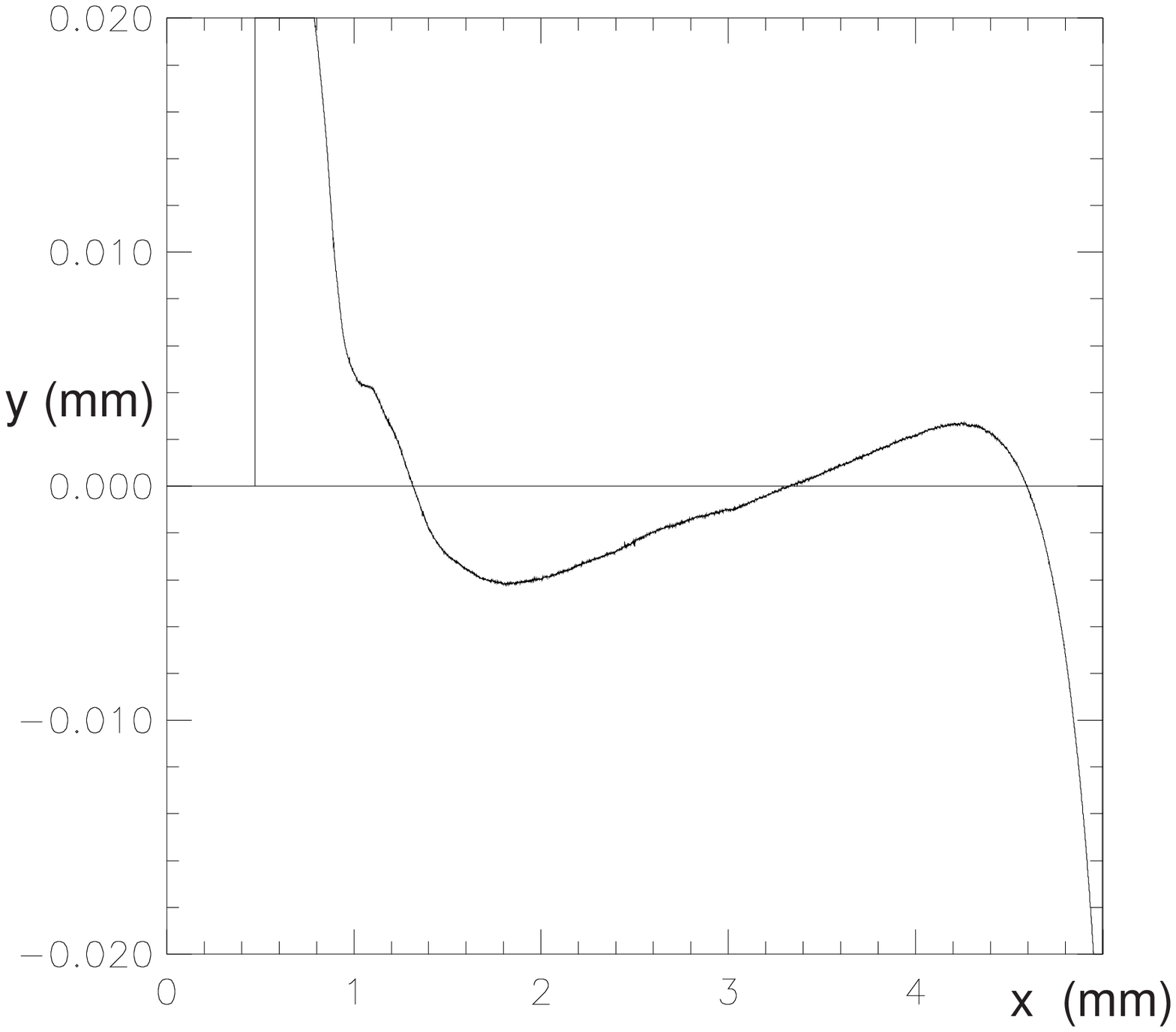,height=2.4in}
\epsfig{figure=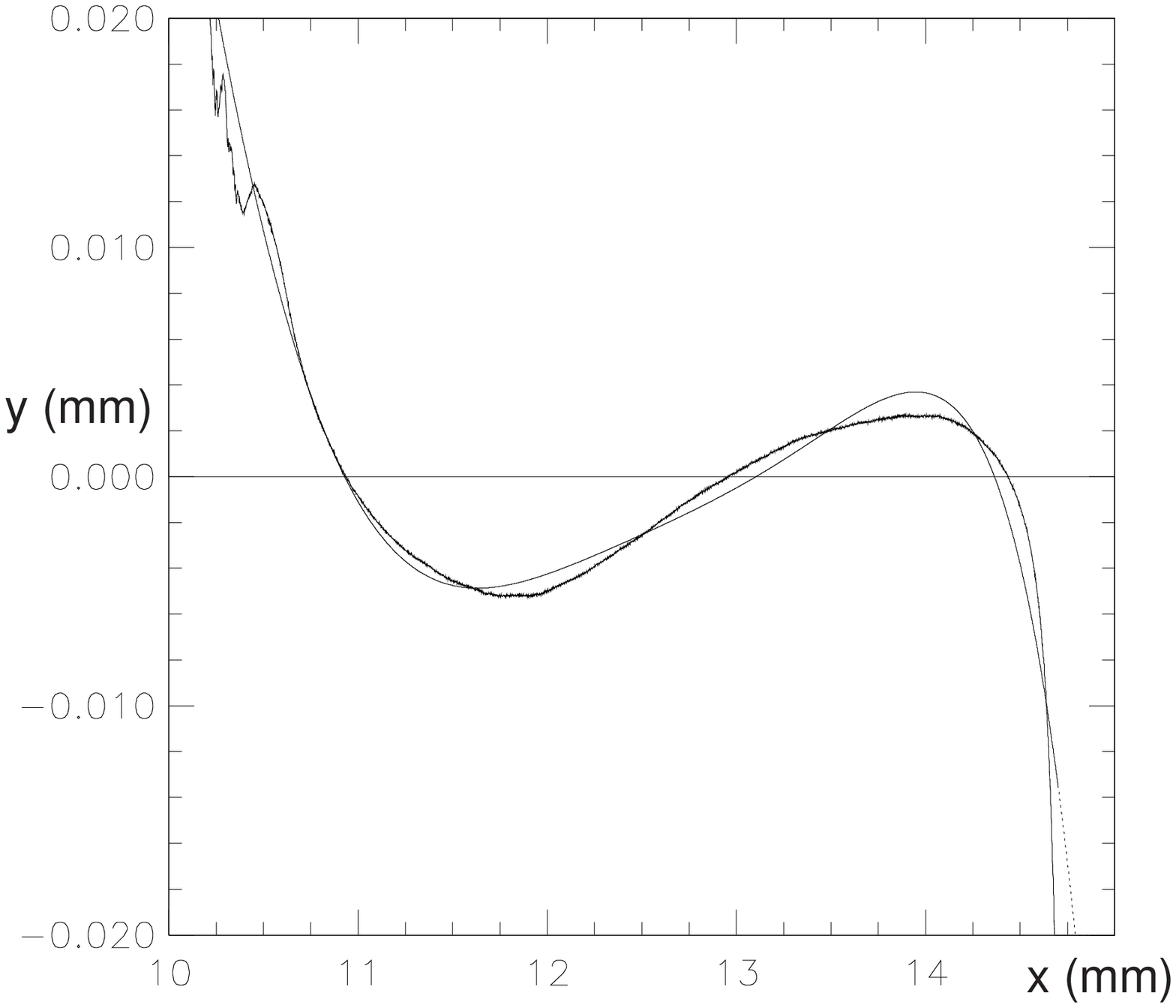,height=2.4in}} 
\vspace{-.3cm} 
\caption{\label{diff/fit/test} 
The deviations from a linear fit for the test beam radiator (left),
and a new radiator with a polnomial fit (right).}
\end{figure}

The actual tooth shapes deviate from strict linearity. Plots of the  
deviation of surface shape with respect to the linear fit are shown in  
Fig.~\ref{diff/fit/test} 
for both radiators. The general features are similar. The non-flat portions are 
caused, on the left side by the roll off in the valleys between the teeth and 
on the right side by the roll off to the tops of teeth. There is almost 1 mm of 
surface with ``large" variations at the bottom and 0.5 mm at the top. Photons 
from these parts of the surface will not be at the proper position. This effect 
 is somewhat ameliorated by the loss of optical transmission especially in the  
 tooth valleys.  
  
\begin{figure}[htb] 
%\vspace{-2.2cm} 
\centerline{\epsfig{figure=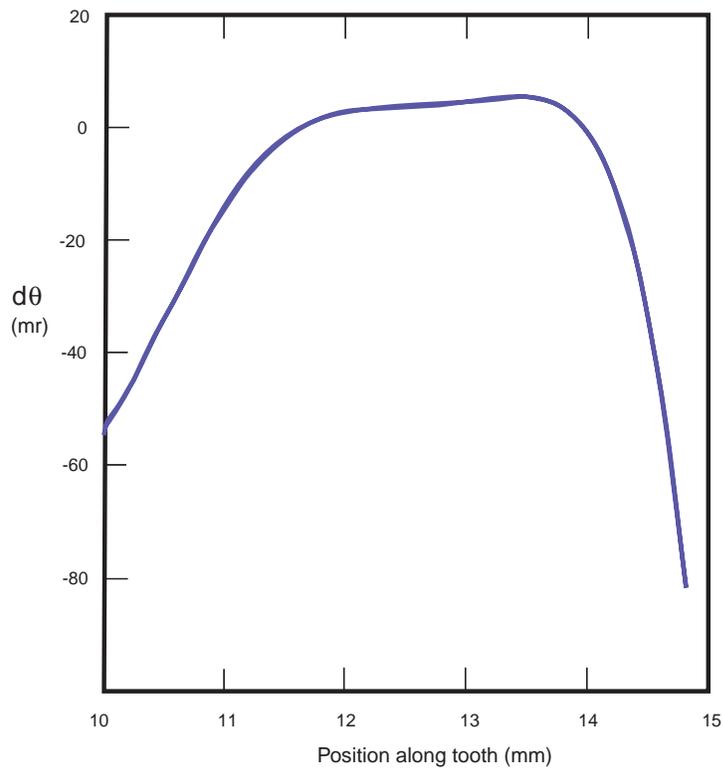,height=4in}} 
%\vspace{-2.7cm} 
\caption{\label{tooth/deriv}The angular deviation of one surface from 
a linear fit based on the fit shown in Fig.~\ref{diff/fit/test} (right).} 
\end{figure}

The plot for the new radiator also shows a polynomial fit to the surface. 
Though the fit isn't perfect it does give a reasonable representation of 
the global aspects of the surface. Using this fit the derivative of the 
deviations from the linear fit is calculated 
and shown in Fig.~\ref{tooth/deriv}.  
  
\subsection{UV Transmission}

\begin{figure}[htb] 
%\vspace{-2.2cm} 
\centerline{\epsfig{figure=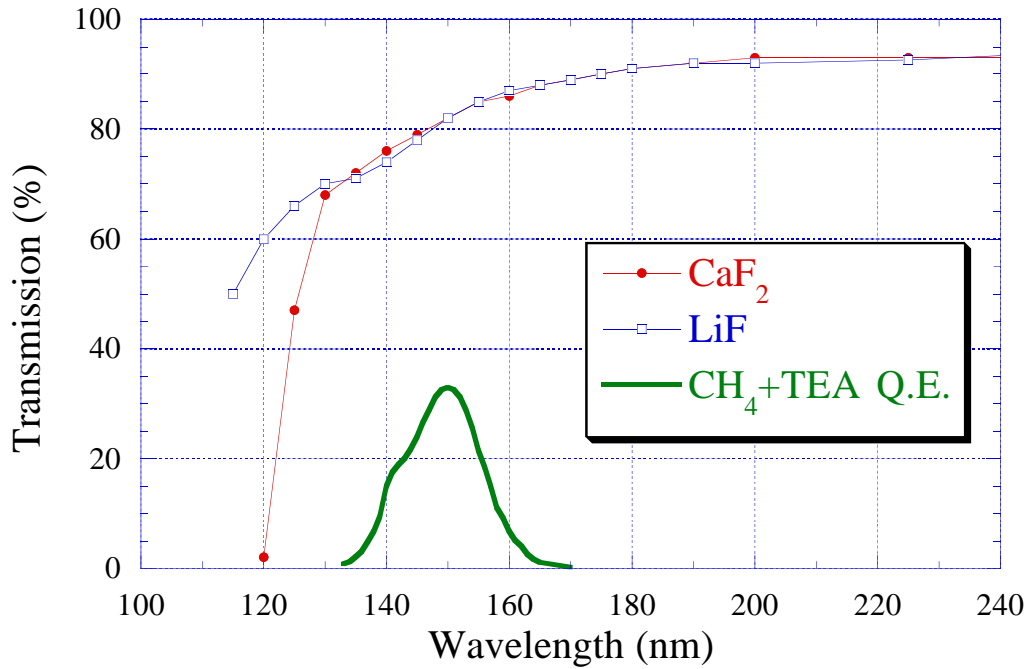,height=3.5in}} 
%\vspace{-2.7cm} 
\caption{\label{plane_trans}The transmission as a function of
wavelength for the plane radiator used in the test beam, the CaF$_2$
windows and the shape of the CH$_4$-TEA quantum efficiency.} 
\end{figure}

The transmission of the radiators, being on average 10 mm thick, is crucial
to observing a significant number of Cherenkov photons. The plane radiators are
polished on both sides and transmission measurements made through the bulk.
On average a photon emerging from a plane radiator will have about half the
absorption probability than indicated from these measurements since it will
traverse only half the bulk material and need to penetrate only one surface.
The measured transmission of a the plane radiator used in this test is shown in
Fig.~\ref{plane_trans}. We also show for comparison the transmission of
an ``average" 2 mm thick CaF$_2$ window, used in the photon detectors and
the shape of the CH$_4$-TEA quantum efficiency curve.

\begin{figure}[htb] 
%\vspace{-2.2cm} 
\centerline{\epsfig{figure=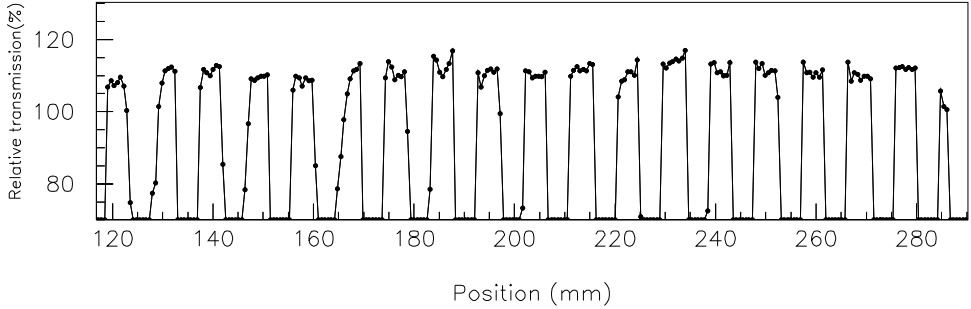,height=1.6in}} 
%\vspace{-2.7cm} 
\caption{\label{trans1} The transmission along a slice of one of the
sawtooth radiators relative to a normally polished prisim at 150 nm. }
\end{figure}

Measurements of the transmission of the sawtooth radiators are made
normalizing to a large prism with the same sawtooth angle that was 
polished using the same technique as the plane radiators. Measurements
are made at 135 nm, 150 nm and 165 nm along three slices of the radiator
and separately for the two sides of a tooth. In Fig.~\ref{trans1} we show
the results of one such measurement at 150 nm.
 
The transmissions are in excess of 100\% showing that this particular radiator
has better transmission than the normally polished prism. The excess could be
due to better bulk transmission, but indicates that excellent polishing
quality has been achieved. The zero transmission bands show the position
of the other side of the tooth which should allow no light through in this
setup. The transmission of that side is determined in a separate scan.
The second sawtooth radiator that we used had 10\% lower transmission
than the first. 

\section{ELECTRONICS} 
 
\begin{figure} [htb]  
\centerline{\epsfxsize 5.5in  \epsffile{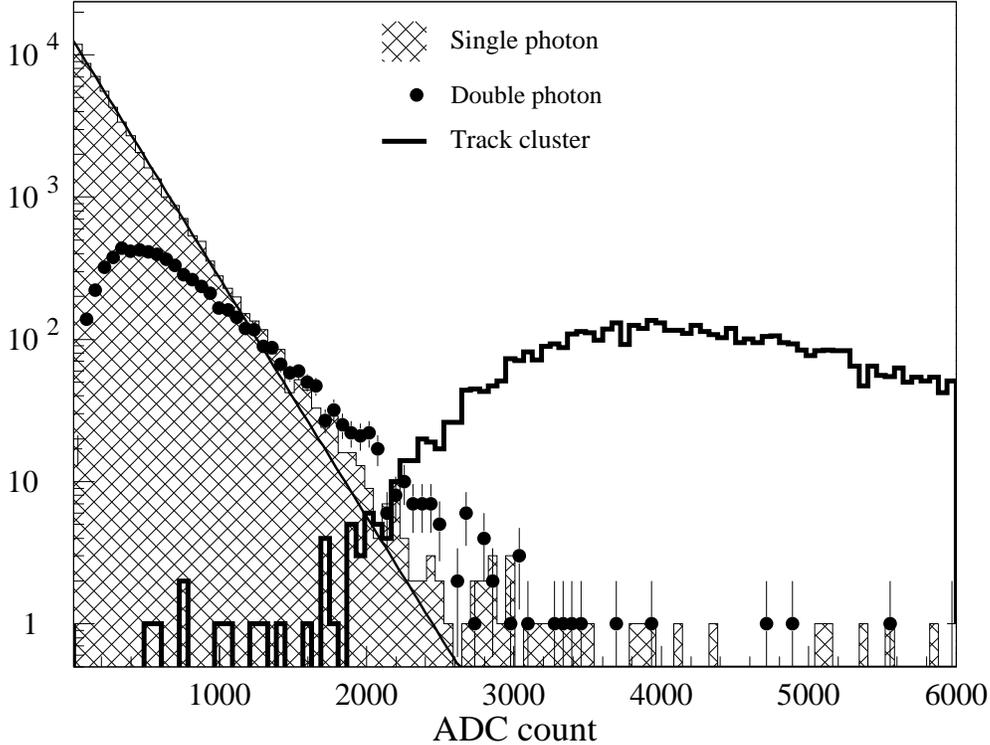}} 
\vspace{-.7cm} 
\caption{\label{pulse} Pulse height distributions from pad clusters containing 
single photons, two photons and charged tracks, from one data run at
a pad gain of $\approx 4\times 10^4$. The line 
shows a fit to an exponential distribution. One ADC 
count corresponds to approximately 200 electrons. The charged track
distribution is affected by electronic saturation.} 
\end{figure}

The position of Cherenkov photons is measured by sensing the induced charge on 
array of 7.5 mm x 8.0 mm cathode pads. Since the pulse height distribution from single 
photons is expected to be exponential \cite{expon}, this requires the use of low noise electronics. 
Pad clusters in the detector can be formed from single Cherenkov photons, 
overlaps of more than one Cherenkov photon or charged tracks. 
In Fig.~\ref{pulse} we show the pulse height distribution for single photons, two 
photons, and charged tracks. (See section(\ref{sec:photoe}) for a discussion 
of how single photon and double-photon pad clusters are defined.) 
  The charged tracks give very large pulse heights because they are traversing 
4.5 mm of the CH$_4$-TEA mixture. The single photon pulse height distribution 
is exponential as expected for moderate gas gain. Also the cluster distribution with two detected  
photons has approximately half the slope of the one photon distribution 
as expected.

 To have as low noise electronics as possible, a 
dedicated VLSI chip, called VA\_RICH, based on a very successful chip  
developed for solid state applications, has been designed and produced for our 
application at IDE AS, Norway. We have fully characterized hundreds of  64 
channel chips, mounted on hybrid circuits. For moderate values of the input 
capacitance $C_{in}$, the equivalent noise charge measured $ENC$ is found to be 
about:  
\begin{equation} ENC = 130 e^- + (9 e^-{\rm /pF}) \times C_{in}~~. \end{equation} 
Its dynamic range is between 450,000 and 900,000 electrons, depending upon 
whether we choose a bias point for the output buffer suitable for signals of  
positive or negative polarity or we shift this bias point to have the maximum 
dynamic range for signals of a single polarity. 
 
In our readout scheme we group 10 chips in a single readout cell communicating 
with data boards  located in VME crates just outside the detector 
cylinder. Chips in the same readout cell share the same cable, which routes
control signals and bias voltages
 from the data boards and output signals to the data boards.
  Two VA\_RICH chips are mounted using wire bonds on one hybrid circuit that
is  attached via two miniature connectors to  the back of the cathode board 
of the photon detector. 
 
The analog output of the VA\_RICH is transmitted to the data boards as a 
differential current, transformed  into a voltage by transimpedance amplifiers 
and digitized by a 12 bit differential ADC. These receivers are  part of  very 
complex data boards which perform several important analog and digital functions. 
Each board  contains 15 digitization circuits and three analog power supply 
sections providing the voltages and currents to  bias the chips, and  
calibration circuitry. The digital component of these boards contains a 
sparsification  circuit, an event buffer, memory to store the pedestal values, 
and the interface to the VME cpu.  
 
While the noise performance in the Fermilab test was acceptable, we 
found that there was relatively large component of coherent noise (about 1000 
electrons) whose source was imperfect grounding of the switched power supplies 
in the DAQ crates. (These will be replaced in the final system.) We achieved 
very good noise performance  by subtracting the coherent noise on each hybrid 
card individually offline. This noise suppression has proven to be so 
successful that  the CLEO data acquisition  group is developing an on line DSP 
implementing this  algorithm, thus making  our system less susceptible to 
coherent pedestal fluctuations.  

\section{BEAM AND TRACKING SYSTEM}

The 800 GeV proton beam was focussed on a thick target (beam dump). Charged
particles which emerged were mostly relativistic muons that covered a large
area. The emerging particles, mostly relativistic muons,  
 had an r.m.s. spread  of 17~cm horizontally and 33~cm 
vertically with its peak 20~cm below the center of the RICH.  
Beam intensities varied from
5 $\mathrm{Hz}/\mathrm{cm}^2$ to
100 $\mathrm{Hz}/\mathrm{cm}^2$. 

%With these intensities we observed 
%significant pileup of events. Therefore runs taken under 
%these conditions were used only for alignment purposes. Data to be used 
%for analysis were taken with a detuned beam and a peak intensity of 
%
 
\begin{figure}[hbt]
       \centerline{\epsfxsize 5in \epsffile{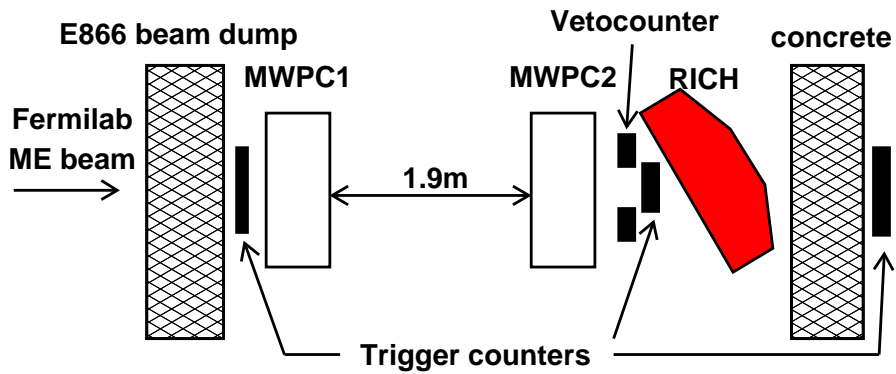} }
       \caption{Test beam setup (not to scale).}
       \label{testbeam_setup}
\end{figure}

 The test beam setup consisted of a 
scintillator hodoscope, two reference MWPCs and
the RICH box as sketched in Fig.~\ref{testbeam_setup}.

We used two identical modules of wire-proportional tracking chambers to define
the position and angle of incident tracks. Each module was 33 cm by 33 cm and
consisted of six planes. The wires were spaced at 5 mm intervals in all planes.
The wire directions were perpendicular to the beam direction and parallel to
the x-axis (horizontal), parallel to the x-axis and displaced by one-half the wire spacing,
parallel to
the y-axis (vertical), parallel to the y-axis and displaced by one-half the wire spacing,
and at +30$^{\circ}$ and -30$^{\circ}$ to the y-axis. The center of the first
module was placed 282 cm from the radiator and the second module 84 cm.

This chamber system had an average layer efficiency of 88\%. The angular
resolution of the charged tracks was $\pm$0.73 mr and the position accuracy
at the radiator was $\pm$1 mm in both x and y directions. The
tracking system inaccuracies are not negligable when compared to the Cherenkov
angle resolution per track. In particular, the position error at the radiator
makes a substantial contribution.

Scintillators in various configurations narrowed the phase space of the
beam particles used for our measurements. The angular dispersions of the 
beam ranged from 3 mr to 6 mr r.m.s., for all the different settings.

\section{TRIGGER} 
The trigger was given by a coincidence of three scintillator counters, 
with the last one behind a wall of $\sim$6 m of concrete to provide a
lower
momentum cut-off of $\sim$3 GeV/c for triggered events ($\beta>0.9994$). 
Additionally, the event was only 
accepted if there was no hit in an anti-counter system surrounding the 
acceptance of the trigger telescope and no additional hit in any of the 
counters in the preceding 100~$\mu$s and following 3~$\mu$s to avoid event 
pile-up due to preamplifier and read-out time constants.

\section{PHOTON DETECTOR HIGH VOLTAGE OPERATION} 
The choice of the operating point of the photon detector wire chambers 
is  driven by the need to optimize the signal-to-noise ratio to 
maximize the number of detected photons while maintaining high-voltage 
stability. The stability in chambers with a gas mixture 
containing TEA is limited by the inherent susceptibility to photon 
feedback. It calls for the use of a powerful quencher gas like 
\(\mathrm{CH}_4\) as the main chamber gas component. In addition the high 
photon absorption capability of \(\mathrm{CH}_4\) limits the sensitive 
spectral region to above 135 nm, thus reducing chromatic errors on the measurement of
the Cherenkov angle.\footnote{The TEA quantum efficiency goes to zero above
165 nm.} Stable operating conditions can be achieved for 
gas gains below $10^6$ \cite{t+j}.

\begin{figure}[hbt]
   \centerline{\epsfxsize 2.7in \epsffile{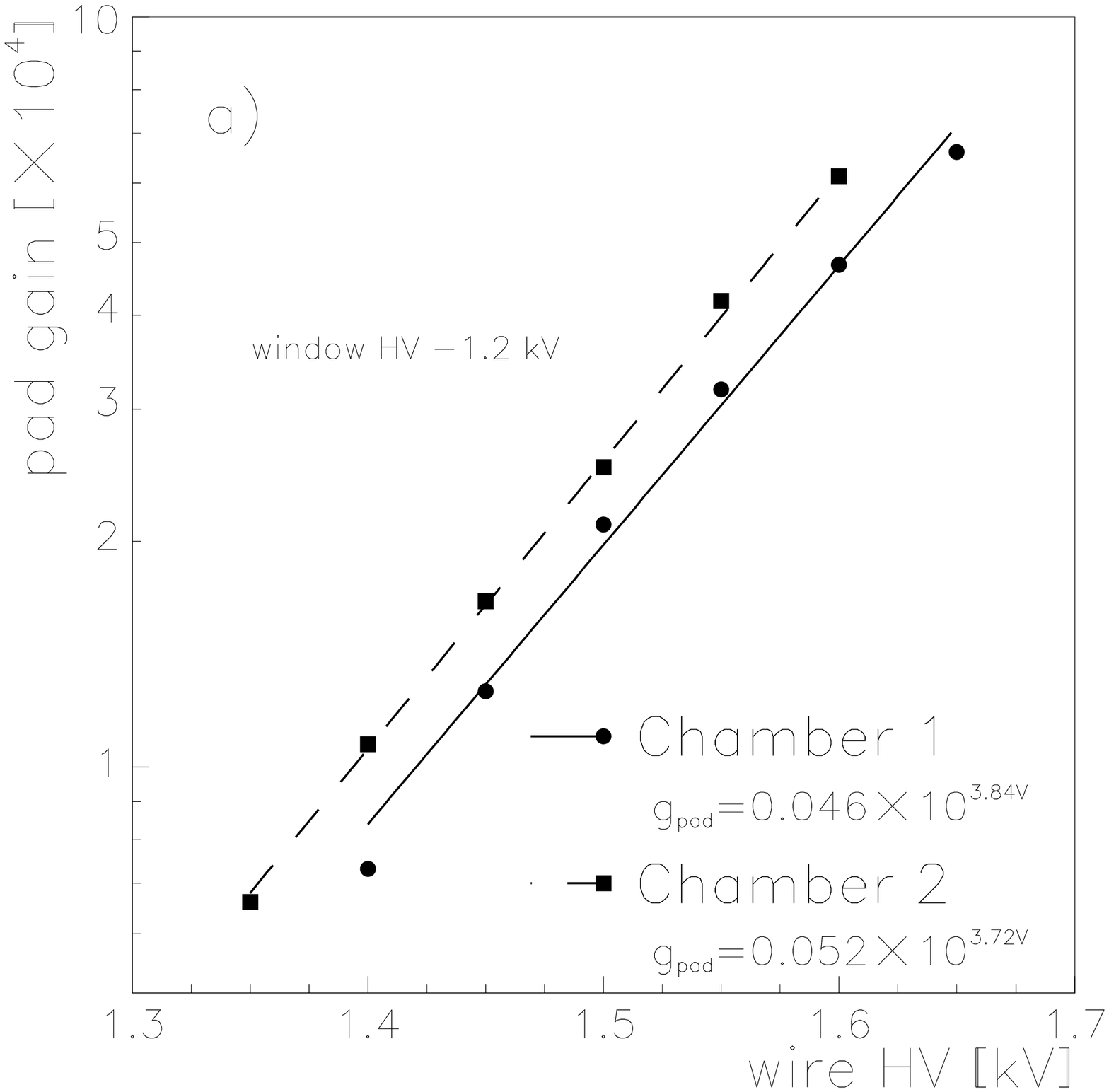} 
            \epsfxsize 2.7in \epsffile{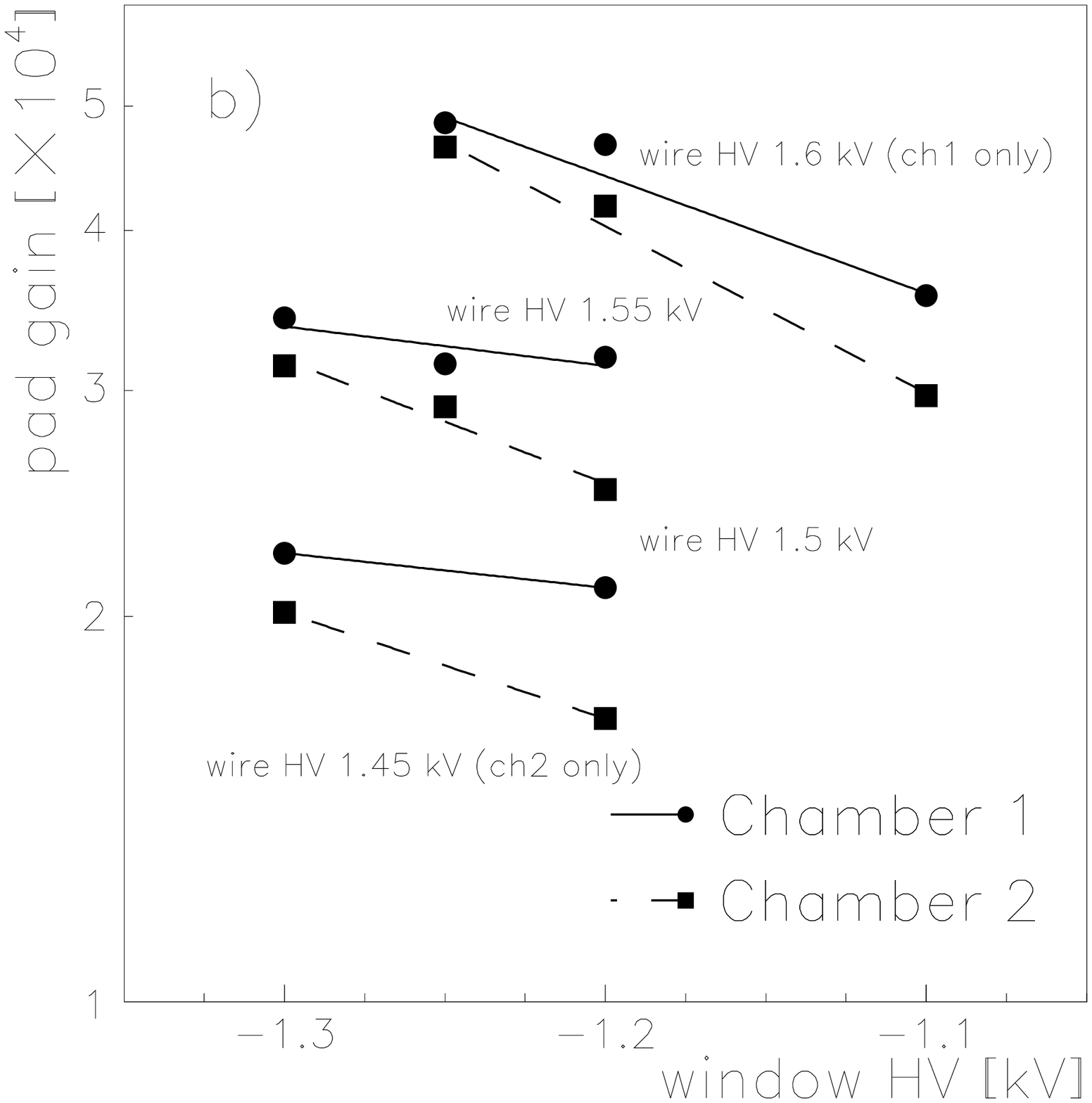}}
   \caption{Pad gain as a function of wire voltage (a) and window voltage (b).
            During each of the the scans the other voltage was kept constant 
            at the level indicated in the plot.}
   \label{hvscan}
\end{figure}

At low gains, $\approx10^{4}-10^{5}$ the gain distribution of our proportional chamber is
a simple exponential~(Fig.~\ref{pulse}). In order to insure that we were
measuring the maximum number of photoelectrons possible, we measured the pad
gain as a function of wire high voltage  and window high  voltage using
the plane radiator.
The results are shown in Fig.~\ref{hvscan}. In both cases the gain increases
exponentially with voltage. Since the wire to cathode spacing is smaller
(1 mm) than the wire to window distance (3.5 mm) it is not surprising that
the dependence on  window voltage is much smaller.

\begin{figure}[hbt]
   \centerline{\epsfxsize 3.4in \epsffile{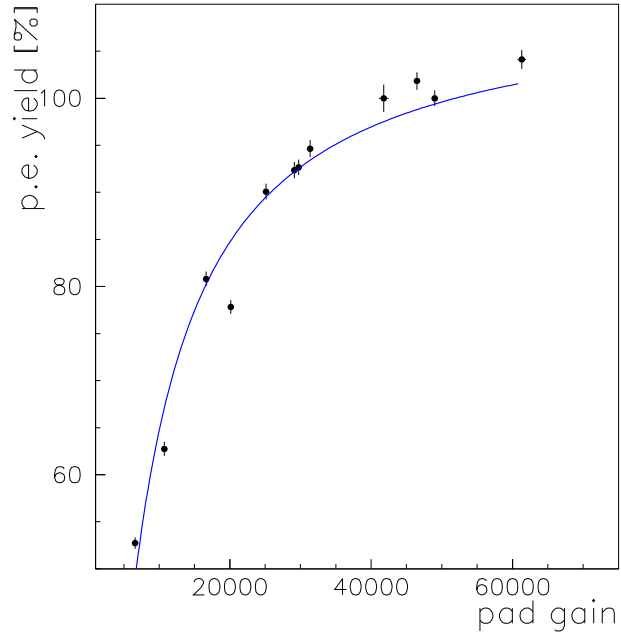} }
   \caption{Detected photoelectron yield as a function of the 
       pad gain. The lower pulseheight threshold is at 2000~$e^{-}$.
The curve is to guide the eye.}
   \label{plateau}
\end{figure}

The exponential gain distribution leads to efficiency loss after a lower 
pulse height threshold is introduced to discriminate electronic noise hits.
For our analysis we set this threshold at $5\sigma_{noise}$, which is
equivalent to 2000 $e^-$.
The operating voltage for the photon detectors during the test beam was 
chosen based on the efficiencies obtained with this pulse height threshold 
during a scan where the pad gain was varied by changing the wire high
voltage. The results are shown in Fig.~\ref{plateau}. Sufficient efficiency
 was achieved when 
the pad gain in the two tested chambers was $>4\times 10^4$. We set the
high voltages to achieve this gain
throughout the test beam measurements. 

The operating stability at the voltage required to achieve such a gain was 
first studied using several 
full-size chamber prototypes and later during chamber construction, when
each chamber was tested for 30~days for high voltage problems.
This procedure allowed for a very stable operation of the finished 
chambers with only one trip exceeding 1~$\mu$A during the four weeks 
test beam period and dark currents usually below 30~nA per group of 24~wires.

%--------------------------------------------------------------------- ----------  
\section{BEAM-TEST RESULTS}  
%---------------------------------------------------------------------  
\subsection{Introduction}\label{sec:photoe} 
\begin{figure}[hbt] 
%\vspace{-4.5cm} 
\centerline{\epsfxsize 3.0in  \epsffile{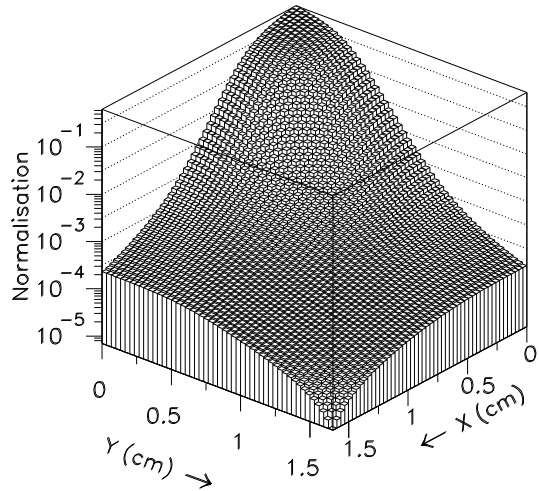}} 
%\vspace{-3.2cm} 
\caption{\label{pulshap} The pulse height seen by a pad centered at
a local coordinate position $(x,~y)$ for a photoelectron
at position (0, 0). The distribution is normalized to 1, for a photoelectron
centered right above a pad.} 
\end{figure}

Photoelectrons  are reconstructed   
by finding topological clusters of pads with each  
pad having a pulse height   
above a 5$\sigma$ noise cut ($\sigma \approx 400 e^-$). 
The location of the photoelectron is determined by first
calculating the naive geometric center-of-gravity and
then apply a correction based on the pulse height   
 profile is shown in Fig.~\ref{pulshap}. This distribution
 was determined using combination of photon data and
 charged track data at normal incidence to the photon
 detectors.  
  At operating voltage,   
pad multiplicities are 2.2 pads per cluster; there is a significant   
overlap of photons. On average there are  1.1 photoelectrons per   
cluster.

From each photoelectron position,   
the original photon trajectory   
is optically traced back   
through all media   
to the center plane of the radiator, and   
the Cherenkov angle is reconstructed.   
Charged tracks through the tracking system are projected to the RICH   
detector where  clusters of pad hits associated with the track are ignored. Other  
charged track clusters are distinguished   
from photon clusters   
by total charge   %(above some threshold)   
and by the number of pads in the cluster. These charged tracks are usually   
out of time with the tracking system, because the RICH preamplifier time
constants are long.

 Data were taken at a variety of incident track angles;   
in the following discussion   
the data taken for tracks at an angle of 30$^{\circ}$ from the normal to the
plane radiator and at normal incidence to   
the sawtooth radiator (0$^\circ$)   
are described in detail. The same analysis techniques were used on  
all data sets.    
 %with teeth parallel to wires, CLEO III orientation of   
%sawtooth radiator   

%----------------  
\begin{figure}[ht] %[hbt]  
  \begin{center}%opt  
       \centerline{\epsfxsize 6.00in \epsffile{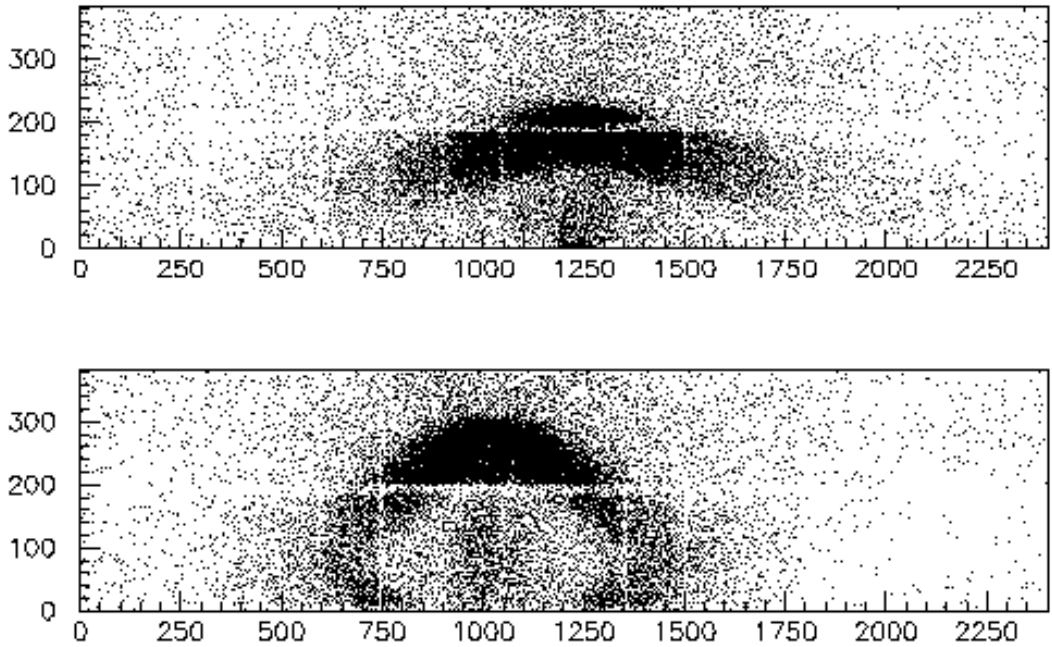}}  
  \end{center}%opt  
\vspace{-7mm} 
\caption{  
Sum of   
11639 Cherenkov images  30$^\circ$ plane radiator dataset (top),   
and   
7804 images in 0$^\circ$ sawtooth radiator dataset (bottom).    
 %Images are seen in both chambers, the bottom one   
Units are mm.   
The bottom chamber contains the beam track, and   
is parallel to the radiator.    
Shadows of structural elements of the photon detectors can be seen.    
(The plane image is at 90$^{\circ}$ to the orientation it will 
have in CLEO.)} 
\label{fig:cum}  
\end{figure}  
%----------------  
  
%%% CUMULATIVE events %%%%%%%%%%%%%%%%%%%%%%%%   
  
Fig.~\ref{fig:cum}   
shows a cumulative event display   
for all ring images in these two sets of data.    
For the plane radiator the one arc Cherenkov ring is   
visible, while for the sawtooth radiator the two arcs   
are visible,   
with the lower one largely outside of the fiducial region of the   
detectors.    
Acceptance is lowered by this image truncation,   
and by mechanical transmission losses from construction elements   
in the detector.   
The acceptance for contained plane radiator images is about 85\%,  
the maximum realistic acceptance for a full RICH system, while   
the acceptance for sawtooth images is   
approximately 50\%\ for this test beam run, because we have only  
two  photon detectors. In the full CLEO setup, the sawtooth   
acceptance is approximately the same as for the plane radiator.    
  
 %sawtooth radiator analysis uses cleo3 orientation   

\subsection{Results from the Plane Radiator at 30$^{\circ}$ Track Incidence}

%%% PLANE radiator results %%%%%%%%%%%%%%%%%%%%%%%%   
  
\begin{figure*}[htb] %[hbt]  
\begin{center}  
\begin{tabular}{ccc}  
       \epsfxsize 5.5in \epsffile{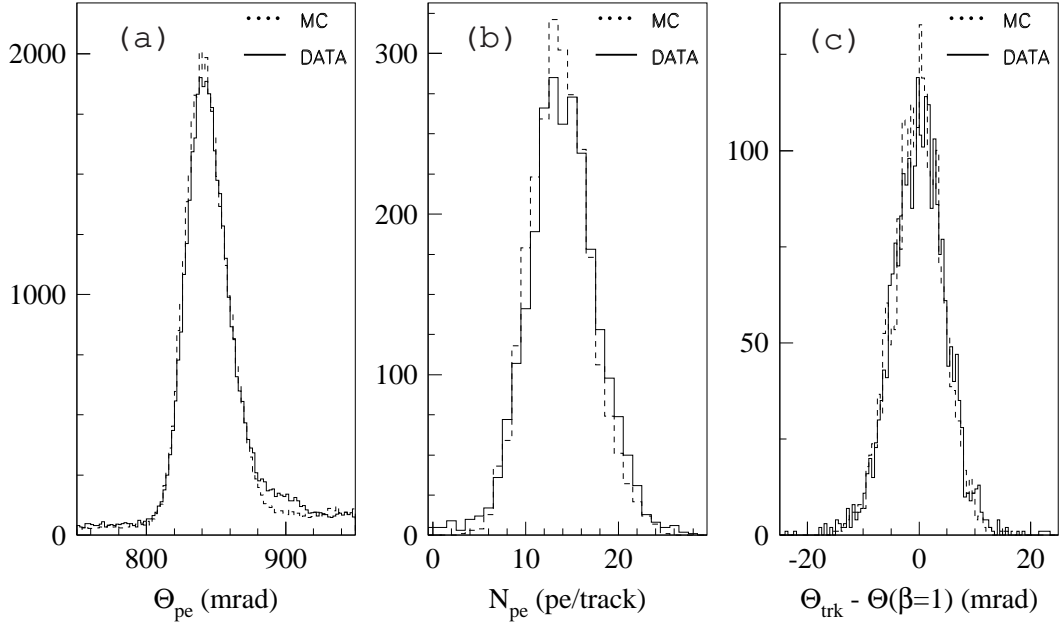}     
\end{tabular}  
\end{center}\vspace{-3mm}  
\caption{  
Plane radiator results at 30$^{\circ}$ track incidence.   
(a) Single photoelectron Cherenkov angle distribution;   
(b) Photoelectron yield (per/track);
(c) Distribution of Cherenkov angle per track   
    shifted by angle of a high momentum muon ($\Theta_{\beta=1}$).   
Solid line is for data, dashed line is for Monte Carlo.   
}  
\label{fig:respl}  
\end{figure*}

Results from the analysis of the 30$^\circ$ plane radiator data set   
are shown   
in Fig.~\ref{fig:respl};   
only images confined to a single detector are used.    
The distribution of Cherenkov angle for single photoelectrons has   
an asymmetric tail and modest background. It is  
 fit with a lineshape similar to that used by for extracting 
 photon signals from electromagnetic calorimeters  
 \cite{CBL}. The functional form is 
\[f(\Theta|\Theta^*,\sigma_{\Theta^*},\alpha,n)= \left( \begin{array}{l} 
  A\cdot{\rm exp}\left[-{1\over 2}\left({{\Theta^*-\Theta}\over \sigma_{\Theta^*}} 
\right)^2\right]~~~~{{\rm for}~\Theta<\Theta^*-\alpha\cdot\sigma_{\Theta^*}}\\ 
 A\cdot{{\left({n\over \alpha}\right)^n e^{-{1\over 2}\alpha^2} 
\over \left({{\Theta^*-\Theta}\over \sigma_{\Theta^*}}+{n\over \alpha}-\alpha\right)^n}} 
~~~~~~~~~~~{{\rm for}~\Theta>\Theta^*-\alpha\cdot\sigma_{\Theta^*}}\\ 
{\rm here}~A^{-1}\equiv \sigma_{\Theta^*}\cdot 
\left[{n\over \alpha}\cdot {1\over {n-1}}e^{-{1\over 2}\alpha^2} 
+\sqrt{\pi\over 2}\left(1+{\rm erf}\left({\alpha\over\sqrt{2}}\right) 
\right)\right]  
\end{array}\right.\] 
\begin{equation} 
\end{equation} 
Here $\Theta$ is the measured angle, $\Theta^*$ is the ``true'' (or most likely)  angle and  
$\sigma_{\Theta^*}$ is the angular resolution. 
 To use this formula, the parameter $n$ is fixed to value of about 5. 
 
\begin{figure}[htb] %[hbt]  
\begin{center}  
\centerline{       \epsfxsize 3.6in \epsffile{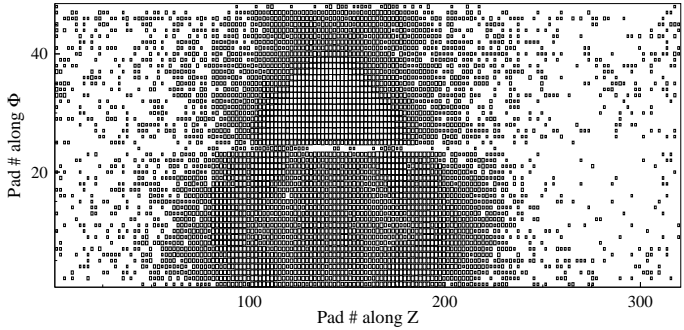}}   
 \end{center}\vspace{-2mm}
\caption{Distribution of photon clusters are shown for a data set using 
the plane 
radiator with normal track incidence, where we do not expect any  
Cherenkov photons due to total internal reflection. Charged track 
clusters have been removed.}  
\label{backsim} 
 \end{figure}

The data in Fig.~\ref{fig:respl} are fit
using this signal shape plus a polynomial background function.   
We find  
a single photoelectron Cherenkov angle resolution   
$\sigma_{\Theta{\rm pe}} = (13.2 \pm 0.05 \pm 0.56)$~mr.   
  This is consistent with the  Monte Carlo estimate of 13.5~mr.   
Errors quoted are first statistical then systematic,   
with the latter taken from two different fitting procedures,   
i.e., two methods of background estimation.    
The background fraction under the peak is 9.2\% .  
This background is   
not electronic noise   
 %(estimated at xxx hit pads per event),   
but rather it is   
principally due to out-of-time hadronic showers   
from the upstream beam dump that acts as our muon filter; 
there will be little or no such background in CLEO III.    
 We tested this hypothesis by taking data with a plane 
 radiator at normal track incidence. In this situation the Cherenkov light 
 cannot escape the radiator due to total internal reflection. The hit pattern 
 on the detector for this situation is shown in Fig.~\ref{backsim}.

The Cherenkov angle per track is found as the arithmetic mean   
of all photoelectrons in an image.  There is an   
image cut of $\pm3\sigma_{\Theta{\rm pe}}$ and   
a systematic alignment correction is applied.    
The resulting distribution of Cherenkov angle per track   
is fit to a Gaussian, and   
gives the angle resolution per track   
$\sigma_{\Theta{\rm trk}} = ( 4.54 \pm 0.02 \pm 0.23 )$~mr,   
which compares favorably with the Monte Carlo estimate of 4.45~mr.   
The systematic error is estimated from   
the variation\footnote{This   
 variation has a number of root causes,   
 each at the few percent level:   
 the expansion volume transparency was monitored   
 to be above 95\%, 	%but not continuously,   
 there are variations in transparency over each radiator face,   
 etc.    
 Hence the systematic error is estimated to be at the 5\%\ level.   
} between different datasets taken at the same track angle,   
which are repeatable 	  
to 5\%.    
  
 %statistical errors are small in all cases   
 %systematic errors dominate in this analysis.   

The photoelectron yield   
$N_{\rm pe} = (13.3 \pm 0.07 \pm 0.36)$ per track   
is extracted from the area under the single photoelectron peak   
followed by background subtraction.    
Again   
systematic errors dominate and are given by different methods of   
background estimation.    
(Here the beam-test Monte Carlo makes no prediction for $N_{\rm pe}$   
but rather uses the measurement as an input parameter.)    
This yield exceeds our benchmark of 12 photoelectrons/track.

%----------------  
%----------------  
  
\begin{figure}[htb]  
  \begin{center} 
       \centerline{\epsfxsize 4.00in \epsffile{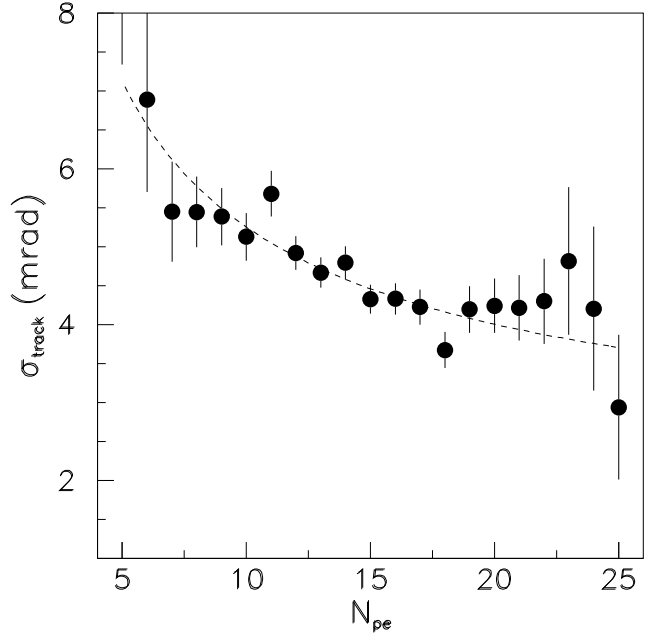}}  
  \end{center}\vspace{-3mm} 
\caption{$\sigma_{track}$ as a function of N$_{pe}$ (not background 
subtracted). The 
curve is a fit to the function $\sqrt{A^2/{\rm N}_{pe}+B^2}$.}  
\label{fig:plane_marina}  
\end{figure}

We can also view the resolution as a function of N$_{pe}$.  
Fig.~\ref{fig:plane_marina} shows the decrease in the track  
resolution as N$_{pe}$ increases. (N$_{pe}$ has not been 
background subtracted; we estimate the background as $\sim$9\%.) 
 The curve is a fit to the  
function $\sqrt{A^2/{\rm N}_{pe}+B^2}$ where, $A=15.2$ and $B=2.1$.  
(These parameters are highly correlated and include all systematic
effects, such as tracking errors.) 
  
%%% SAWTOOTH radiator results %%%%%%%%%%%%%%%%%%%%%%%%   
 \subsection {Sawtooth Radiator Results at Normal Track Incidence} 
\begin{figure}[hbt]  
\begin{center}  
\begin{tabular}{ccc}  
       \epsfxsize 5.5in \epsffile{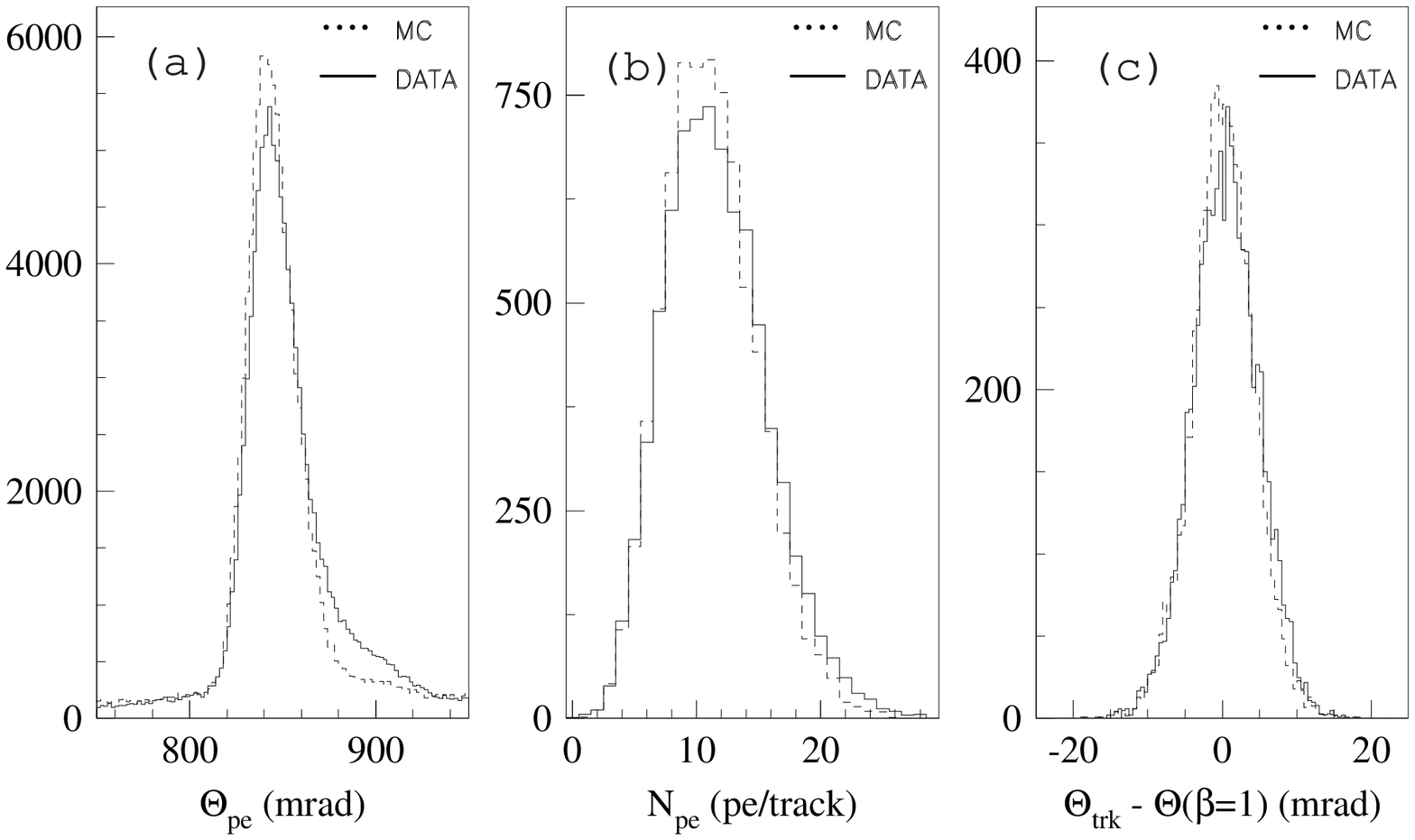} 
\end{tabular}  
\end{center}  
\caption{  
Sawtooth radiator results for $0^{\circ}$ track incidence.   
(a) Single photoelectron Cherenkov angle distribution;   
 (b) Photoelectron yield (per track);
 (c) Distribution of Cherenkov angle per track   
    shifted by angle of a high momentum muon ($\Theta_{\beta=1}$).   
Solid line is for data, dashed line is for Monte Carlo.   
}  
\label{fig:resst}  
\end{figure}

Similar analysis for the 0$^\circ$ sawtooth radiator dataset,   
cf.\ Fig.~\ref{fig:resst},   
gives   
a single photoelectron Cherenkov angle resolution   
$\sigma_{\Theta{\rm pe}} = (11.7 \pm 0.03 \pm 0.42)$~mr   
  (compared with 11.1~mr from Monte Carlo),   
an angle resolution per track   of
$\sigma_{\Theta{\rm trk}} = ( 4.49 \pm 0.01 \pm 0.22 )$~mr   
(4.02~mr from Monte Carlo),   
and   
a photoelectron yield   
$N_{\rm pe} = (10.4 \pm 0.04 \pm 1.0)$ per track. 
The background fraction of 12.0\% has
been subtracted.   
Adjusted for full 85\%\ geometric acceptance,   
$N_{\rm pe}$ becomes 18.8 per track.

 %Note the enhanced tail at higher angles.    
 %mc sawtooth `defect' plots  

%----------------  
%----------------  
 \subsection{Summary  and Discussion of Results} 
 
The average observed photon yields for the plane and sawtooth radiators at
various incident angles are 
given in Table~\ref{table:ngamma}. 
The plane data is for full acceptance, while the sawtooth data is given both 
for observed and extrapolated yields. 
The statistical errors are negligible in comparison to the systematic errors 
which we estimate to be $\pm$5\%. 
 
\begin{table}[th]\centering\caption{Cherenkov Photon Yields } 
\label{table:ngamma} 
\vspace*{2mm} 
\begin{tabular}{lrrrrr}\hline\hline 
Radiator & 0$^\circ$ & 10$^\circ$ & 20$^\circ$ & 30$^\circ$ &
40$^\circ$\\\hline 
plane & & & 14.1 & 13.3 & 12.3\\ 
sawtooth (observed) & 10.4 & 10.7 & 10.1 & 9.7 & 9.5\\ 
sawtooth (full acceptance) & 18.8 & 19.2 & 17.9 & 16.3 & 15.5\\ 
\hline\hline 
\multicolumn{6}{l}{Systematic errors are $\pm$5\%, statistical errors are much smaller.} 
\end{tabular}\end{table} 
  
%%% SUMMARY plot %%%%%%%%%%%%%%%%%%%%%%%%   
  
\begin{figure}     [hbt]  
  \begin{center}%opt  
       \centerline{\epsfxsize 5.500in \epsffile{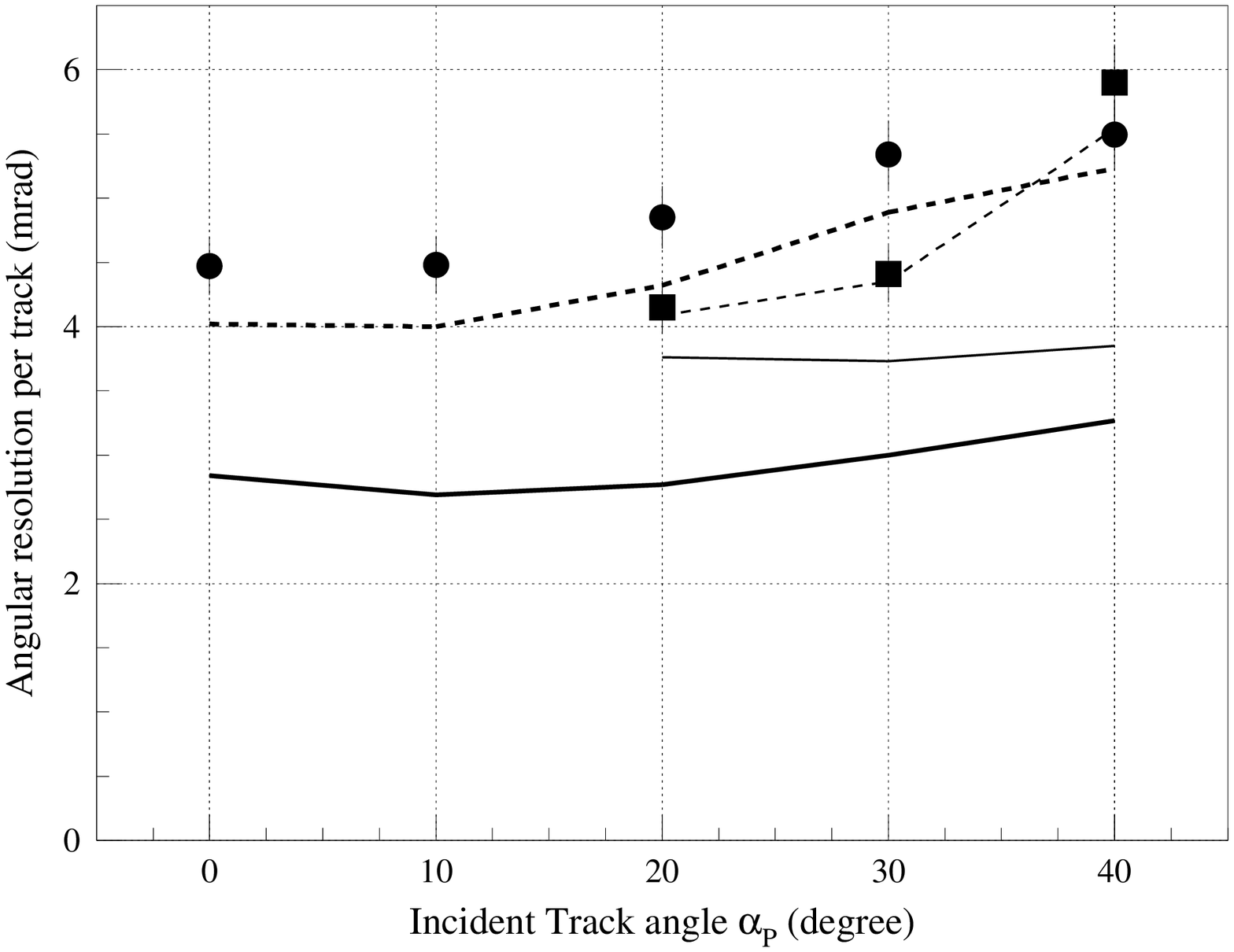}}  
  \end{center}%opt
\vspace{-2mm}  
\caption{  
Summary of CLEO III RICH beam-test results.   
Squares indicate plane radiator results, and   
circles sawtooth radiator results.   
The filled symbols represent results from beam-test data,   
dashed curves results from beam-test Monte Carlo,   
and   
solid curves results from the ``full acceptance'' extrapolation, which
also hypothesises no background or tracking errors.    
}  
\label{fig:ressum}  
\end{figure}

Fig.~\ref{fig:ressum} provides a summary of angular resolution results   
from all datasets at all incident angles.   
The measured Cherenkov angle resolution per track from the plane   
radiator data   
(denoted by squares)   
increases as a function of the incident track angle   
due to the increase   
in emission-point error.\footnote{The   
  Cherenkov angle resolution per track is dominated   
  by chromatic and emission-point errors \cite{t+j}.   
  The chromatic error is larger at small incident track angles,   
  but they become comparable at large incident track angles.   
}   
The beam-test Monte Carlo simulation gives the light dashed curve   
in Fig.~\ref{fig:ressum},   
which represents the data well.    
 %for these angles.    

%----------------  
%----------------    
  
The per track resolution, e.g. 4.54~mr at 30$^\circ$, is   
somewhat larger   
than the nave extrapolation from angular resolution per photon and the number of observed photons,   
i.e. $13.2\, {\rm mr}/\sqrt{13.3}=3.62\, {\rm mr}$.   
 %statistical argument that all photons from a track are uncorrelated   
 %  
Monte Carlo studies indicate that   
the sources of the increased resolution   
are the MWPC tracking errors (the principal cause, $\pm$2.3~mr at   
30$^\circ$)   
and the beam background ($\pm$1.2~mr at 30$^\circ$).    
The tracking errors in this experiment do not change with incident angle
to the radiator, since the angle is changed by rotating the RICH box.
However, the emission-point error is effectively increased more with
larger incident angles to the radiator,  
due to an incorrect track impact point on the radiator face.
In CLEO III the background will be much reduced,   
and the tracking error contribution will be smaller,  
though still significant.   
  
  We expect that the sawtooth radiator   
resolution will  not be degraded by the tracking errors \cite{alex} in CLEO,
 while the plane   
radiator resolution is worsened and the degradation increases with dip   
angle. The tracking system has its poorest   
resolution in the direction    along the electron beam. The sawtooth   
images are symmetric in this direction, and thus are less sensitive to   
the tracking errors than the plane radiator images. The tracking   
errors cause poorer  resolution at larger angles.

The measured track resolutions for the sawtooth radiators
are higher than predicted by 
 the beam-test Monte Carlo as  
shown by the heavy dashed curve in Fig.~\ref{fig:ressum}.   
 Our simulation shows that
this 
discrepancy is not due to the variation in tooth angle
or the actual shape of the radiator surfaces.   
The relatively large values of the track resolution,
 above 4 mr, is due to
the limited acceptance, caused by having only
two photon detectors.
     
In order to estimate the   
ultimate performance of this RICH,   
 %performance of the RICH expected in the CLEO III configuration,   
an extrapolation was made based on the beam-test Monte Carlo.   
 %  
 %As stated,   
The background and tracking errors   
 %will not be so egregious in CLEO III as they were in the   
are associated only with our beam test, so both   
were removed from the simulation for this study.    
The resulting photoelectron yield was then   
corrected for the geometric acceptance of the beam-test setup   
and scaled to ``full acceptance'',   
defined as 85\%\ of the solid angle covered   
by a cylindrical RICH. This correction was applied to the
sawtooth data only, as the plane radiator data are already
at full acceptance.
 %by the full 30 module CLEO III RICH.   
 %  
The result of this                     extrapolation for the   
per track resolution for the plane radiator   
is shown as the light solid curve in Fig.~\ref{fig:ressum},   
which is flat in track angle and   
below our benchmark of 4 mr for CLEO III.

         For the sawtooth a large acceptance correction
needs to be made.
The 
Geometric acceptance                            
        is approximately 50\%\ for all track angles.   
By naive statistical calculation this   
increases the per track resolution by 35\%.   
Monte Carlo studies show that   
tracking errors are the next largest contribution   
(e.g. 1.9~mr at 0$^\circ$),   
and are exacerbated in this configuration because one of the arcs in   
the   
image is out of the detector fiducial.    
The beam background   
is approximately constant for all track angles   
(e.g. 1.3~mr at 0$^\circ$).

          The full acceptance extrapolation   
for the per track resolution   
for the sawtooth radiator, is shown as  
the heavy solid curve in Fig.~\ref{fig:ressum}. 
Here the muon tracking errors and the 
background have been removed and the photon acceptance
of the sawtooth scaled to that of the plane radiator.
 The expected CLEO III RICH performance will fall somewhere   
between the beam-test Monte Carlo curve   
and the ``full acceptance'' curve.  
  
\begin{figure}[t] %[hbt]  
  \begin{center}%opt  
       \centerline{\epsfxsize 4in   
\epsffile{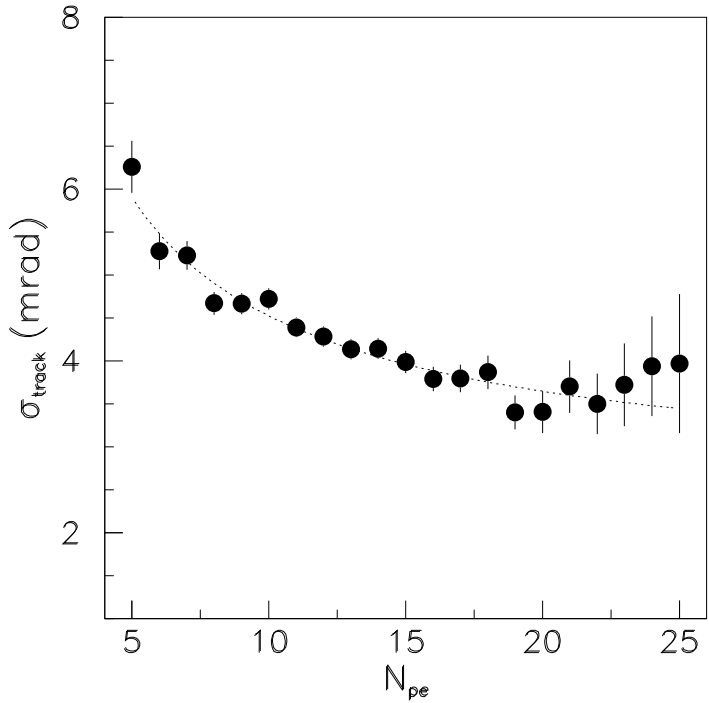}}  
  \end{center}%opt
\vspace{-7mm}  
\caption{The Cherenkov angle resolution per track   
as a function of photoelectron yield   
for sawtooth data at 0$^\circ$ track incidence.    
The curve is a fit to $\sqrt{A^2/{\rm N}_{pe}+B^2}$, where $A$=12.0 and 
$B$= 2.48; these values are highly correlated and affected by the
tracking errors and backgrounds.}  
\label{fig:mplot}  
\end{figure}

 Instead of extrapolating the resolution at the larger
 projected photoelectron yields, we can view the resolution per track already
 achieved in this data as a function of photoelectron yield. 
Fig.~\ref{fig:mplot} shows the   
Cherenkov resolution per track for the 0$^\circ$ sawtooth dataset   
as a function of photoelectron yield.    
For the expected yield of 18.8 per track, the Cherenkov resolution
is 3.5~mr. This number which contains all experimental effects
fits nicely between the best allowable resolution of 3 mr and below
the measured average 4.5 mr.	  
Since this curve is derived from the data   
it automatically takes into account   
statistical and systematic effects.

%----------------  
%----------------  

\begin{figure}[htb] %[hbt]  
  \begin{center}%opt  
       \centerline{\epsfxsize 5.50in \epsffile{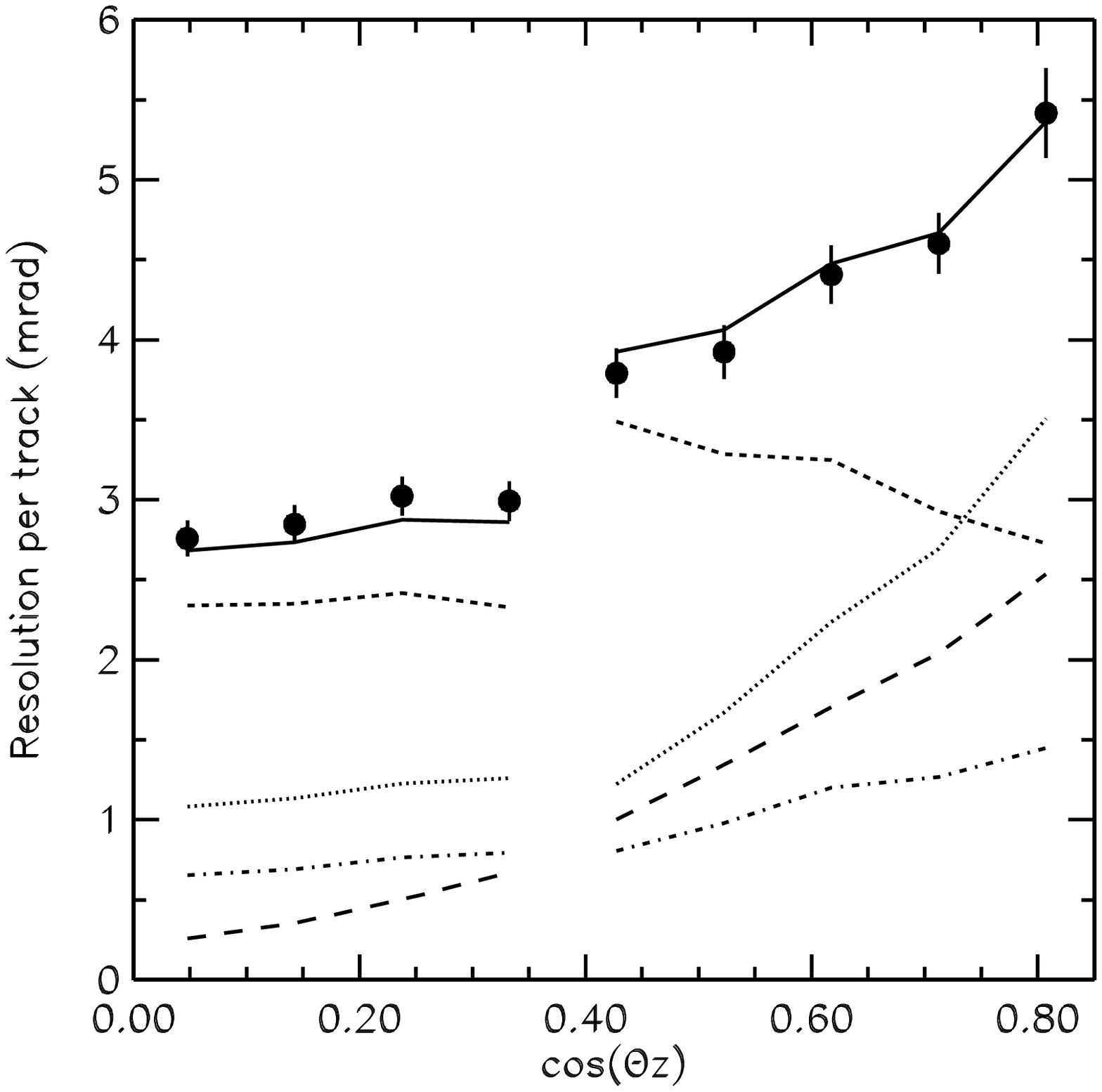}}  
  \end{center}%opt  
 \vspace{-2cm}  
\caption{  
Predictions for CLEO III RICH resolution as a function of 
the cosine of the dip angle, $\Theta z$. The solid points show the 
total $\sigma$. The lines give the contributions of various 
individual error sources:  
tracking error (long dash), photon position determination (dot-dashed), 
lack of knowledge of photon emission point (dotted) and chromatic 
(short dash). The solid line is the sum of the individual components 
added in quadrature and its agreement with the points serves as a  
cross-check. Points for $\cos(\Theta z)<0.40$ represent sawtooth 
radiators, while the other points are for plane radiators.} 
\label{Oct20_k_trk}  
\end{figure}

  In Fig.~\ref{Oct20_k_trk} we show the expectations of the resolution in CLEO III,   
where we have included the CLEO III tracking errors. The largest source of 
error for the sawtooth radiator is the chromatic dispersion, which is 
even larger for the plane radiator. However, the plane radiator also has 
large contributions from the emission point error (not knowing where in 
the radiator the photon was emitted) and the tracking error. Averaging 
over the solid angle the Cherenkov angular resolution will be approximately 
4 mr, which is adequate to separate kaons from pions in the two-body $B$ 
decays to $K\pi$ or $\pi\pi$, which have an average difference in Cherenkov 
angle of 14.4 mr. 
 
%---------------------------------------------------------------------  
 
\section{CONCLUSIONS} 		%STATUS AND SUMMARY}    
%---------------------------------------------------------------------  
 
 %  
A beam test of the first two sectors of the CLEO III RICH Detector  
has been successfully carried out. The resolution is better at 
smaller incident track
    angles and better for sawtooth radiators. The detector 
operated in a robust manner and will make a useful particle identifier 
for CLEO III.

For  plane radiators we measure a full acceptance yield
of between 12-14 photoelectrons, and Cherenkov angular
resolutions of 4-5 mr, depending on incident angles between
20$^{\circ}$ and 40$^{\circ}$.
Our results are consistent with Monte Carlo expectations taking into
account tracking errors and backgrounds.  

For sawtooth radiators we must extrapolate to full acceptance.
We expect between 18-19 photoelectrons between 0$^{\circ}$ and 20$^{\circ}$
incident track angles.
The Monte Carlo under estimates somewhat the measured Cherenkov angular 
resolution. However, even including an anomalous contribution of 1.4 mr
in quadrature with our Monte Carlo expectations of  2.8 mr gives a projected
CLEO III resolution of 3.1 mr. 

Since the difference in $\pi/K$ Cherenkov angles is 14.4 mr in the highest
momentum range of $B$ decays, we expect that this detector will be useful
indeed.
  
 %  

%---------------------------------------------------------------------  
 
\section*{ACKNOWLEDGEMENTS}  
%---------------------------------------------------------------------  
 
  %  
We would like to thank Fermilab for providing us with   
  %the necessary infrastructure and   
the dedicated beam time for our test,   
the Computing Division for its excellent assistance,   
and our colleagues from E866   
for their hospitality in the beamline. Especially 
helpful were Chuck Brown, Joel Butler, and Ruth Pordes.
We appreciate the initial guidance and advice from
Jacques ~S\'{e}guinot and Tom Ypsilantis.
Franz Muheim and Stephen Playfer were important 
participants in the early stages of this work. 
We also thank the U. S. National Science Foundation and
the Dept. of Energy for support.

%---------------------------------------------------------------------  
----------  
\end{document}